\documentclass{ws-ijmpd}
\usepackage[super,compress]{cite}
\usepackage[]{graphicx}
\usepackage{sidecap}
\usepackage{amssymb}
\usepackage[latin1]{inputenc}
\usepackage{amsmath}
\usepackage{amsfonts}
\usepackage{amssymb}
\usepackage{txfonts}
\usepackage{pxfonts}
\usepackage{epstopdf}
\usepackage{float}
\def\lsim{\lower.5ex\hbox{$\; \buildrel < \over \sim \;$}}
\def\gsim{\lower.5ex\hbox{$\; \buildrel > \over \sim \;$}}
\renewcommand{\d}{\mathrm{d}}

\begin{document}

\markboth{Tarafdar \& Das}
{Acoustic surface gravity in Schwarzschild metric}

\title{DEPENDENCE OF ACOUSTIC SURFACE GRAVITY ON GEOMETRIC CONFIGURATION 
OF MATTER FOR AXIALLY SYMMETRIC BACKGROUND FLOWS IN THE
SCHWARZSCHILD METRIC}

\author{PRATIK TARAFDAR}

\address{S. N. Bose National Centre for Basic Sciences, Block JD, Sector III, Salt Lake, Kolkata, India. \\
pratik.tarafdar@bose.res.in}

\author{TAPAS K. DAS}

\address{Harish Chandra Research Institute, Chhatnag Road, Jhunsi, Allahabad 211019, UP, India. \\
tapas@hri.res.in}

\maketitle

\begin{history}
\received{Day Month Year}
\revised{Day Month Year}
\end{history}

\begin{abstract}
\noindent
In black hole evaporation process, the mass of the hole anti-correlates
with the Hawking temperature. This indicates that
the smaller holes have higher surface gravity. For analogue Hawking effects, 
however, the acoustic surface gravity is determined by the local values 
of the dynamical velocity of the stationary background fluid flow and the 
speed of propagation of the characteristic perturbation embedded in the 
background fluid, as well as by their space derivatives evaluated 
along the direction normal to the acoustic horizon, respectively. The mass 
of the analogue system - whether classical or quantum - does not directly
contribute to extremise the value of the associated acoustic surface gravity.
For general relativistic axially symmetric background fluid flow in the 
Schwarzschild metric, we show that the initial boundary conditions 
describing such accretion influence the 
maximization scheme of the acoustic surface gravity and associated analogue 
temperature.
Aforementioned background flow 
onto black holes can assume three distinct geometric 
configurations. Identical set of initial boundary conditions can lead to 
entirely different phase-space behavior of the stationary flow solutions,
as well as the salient features of the associated relativistic acoustic geometry. 
This implies that it is imperative to investigate how the measure of the 
acoustic surface gravity 
corresponding to the accreting black holes gets influenced by the
geometric configuration of the inflow described 
by various thermodynamic equations of state. 
Such investigation is useful to study the effect of 
Einstenian gravity on the non-conventional
classical features as observed in Hawking like effect in a dispersive medium 
in the limit of a strong dispersion relation. 
\end{abstract}


\ccode{PACS numbers: 04.40.Dg, 04.70.Dy, 95.30.Sf}

\section{Introduction}
\label{section1}
\noindent
Black hole analogues are fluid dynamical analogue of the black hole space time 
as perceived in the general theory of relativity 
\cite{unr81,unr95,vis98,bil99,nov02,car05,bar05,us07}.
Such analogue systems may be realized by studying the propagation of 
small amplitude linear perturbation 
through a dissipationless, irrotational, barotropic transonic fluid.
Contemporary research in the field of analogue gravity phenomena has gained widespread 
currency since it opens up the possibility of understanding the salient features 
of the horizon related effects through experimentally realizable physical 
configurations within the laboratory set up. 

Conventional works in this direction,
however, concentrate on systems not directly subjected to the gravitational force. 
Gravity like effects are manifested as emergent phenomena. 
In such cases, only the Hawking like effects can be studied and no direct connection 
can be made to such effects with the general relativistic Hawking effects since such 
non gravitating systems do not include any source of strong gravity capable of 
producing the Hawking radiation. 

To explore whether (and how) the emergent gravity phenomena
may be observed in a physical system which itself is under the influence of a strong 
gravitational field, a series of recent works describe
how the acoustic geometry may be realized for stationary, spherically and axially symmetric hydrodynamic 
flow onto astrophysical black holes 
\cite{das04,das05,abd06,das07,mac08,mac09,nag12,pmdc12}. 

Accreting black holes represent
systems which simultaneously contain gravitational as well as acoustic horizons and 
are shown to be natural examples of large scale classical analogue systems found in the 
universe. This allows us to study the influence of the original 
background black hole space time metric on the embedded perturbative acoustic 
metric. 

Axisymmetric, general relativistic, low angular momentum, inviscid hydrodynamic  
accretion onto non-rotating astrophysical black holes can be studied 
for three different geometric configurations of matter -- disc accretion with constant flow thickness 
(hereafter constant height flow), quasi-spherical accretion in 
conical configuration (hereafter conical flow), and for 
axisymmetric flow maintained in the hydrostatic equilibrium along the 
vertical direction (hereafter vertical equilibrium flow). Details
about such geometric configurations can further be found in \cite{nag12} and in section 4
of \cite{bcdn12}.
For these three geometric configurations,
the nature of the sonic geometry embedded within the infalling material 
has recently been studied 
for accretion processes
under the influence of the generalized post-Newtonian pseudo-Schwarzschild potentials
\cite{bcdn12}. 

In our present work, we would like to extend such calculations 
on a more formal foundation. We shall study the properties of the 
sonic geometry for general relativistic axisymmetric accretion 
(resulting in the existence of a curved background geometry for the 
stationary fluid configuration) onto a Schwarzschild black hole 
for three different geometric configurations of non self gravitating 
matter and for each configuration, two different thermodynamic equations of state. 
We would like to understand how crucial is the role of 
the relativistic gravitation as well as the geometric configuration of the 
background stationary flow (subjected to that gravitational field) in 
determining the essential features of the analogue gravity phenomena. 
We thus intend to demonstrate how the estimation of the 
acoustic surface gravity $\kappa$ gets influenced by the geometric configuration of matter 
for {\it general relativistic background matter flow} onto a Schwarzschild 
black hole. 

For canonical Hawking effect in connection to a 
Schwarzschild black hole, the Hawking temperature $T_{\rm AH}{\propto}\frac{1}{M_{\rm BH}}$ 
({${M_{\rm BH}}$ being the mass of the Hawking radiating black hole). This
indicates that
one requires a 
black hole of reasonably small mass -- e.g., a primordial black hole of
cosmological origin -- to 
maximize the observable Hawking effect. 
The extremisation of the observable Hawking effect can thus be parameterized 
by the mass of the black hole only. One can not have such a straight forward (anti) correlation available for 
the analogue temperature with the mass parameter of the system to 
conclude that the acoustic black hole of microscopic dimension will 
indeed produce a larger analogue temperature. Extremisation of such
temperature as well as $\kappa$
depends on various initial boundary conditions determining the 
background stationary states of the system under consideration. 

\section{Acoustic surface gravity ($\kappa$) for accreting black hole systems}
\label{section2}
\noindent
For a stationary 
flow configuration, the acoustic horizon is the surface 
defined by the equation \begin{equation}
u_{\perp}^2-c_s^2=0 ,
\label{anal2}
\end{equation}
where $c_s$ is position dependent sound speed
(the speed of propagation of the perturbation under 
consideration in general), the bulk flow velocity $u_\perp$ is 
measured along the direction normal to the 
acoustic horizon. 

For any general flow model in Minkowskian 
spacetime,
the acoustic surface gravity $\kappa$
measured at the acoustic horizon 
$r_h$      
can be obtained as \cite{unr81,vis98}
\begin{equation}
\kappa~{\propto}
\left[c_s\frac{\partial}{\partial{\eta}}
\left(c_s-u_\perp\right)\right]_{ r_h} ,
\label{anal3}
\end{equation}
where space gradient 
$\partial/\partial{\eta}$ is taken along the normal to the acoustic horizon.
The subscript $r_h$ indicates that the quantities under consideration  
have been evaluated on acoustic horizon.

Corresponding relativistic 
generalization of the expression of the acoustic surface gravity is expressed as \cite{bil99,abd06,das07}
\begin{equation}
\kappa=\left[\frac{\sqrt{{\chi^\mu}{\chi_\mu}}}{\left(1-{c_s}^2\right)}
\frac{\partial}{\partial{\eta}}\left(u_\perp-{c_s}\right)\right]_
{ r_h} ,
\label{anal4}
\end{equation}
where $\chi^\mu$ is the Killing field which is null on the corresponding
acoustic horizon.
In subsequent sections, we will show that 
the explicit expression for the norm of $\chi^\mu$ can be evaluated in terms of 
the values of the background metric 
elements evaluated on the acoustic 
horizon and on certain flow parameters.

One needs
to calculate the location of the acoustic horizon $r_h$ for a stationary 
configuration, as well as to evaluate the 
expression for the normal bulk flow velocity $u_\perp$ and the speed of the 
propagation of the acoustic perturbation $c_s$ along with their space 
gradients normal to the acoustic horizon to compute the value of $\kappa$.

For the background stationary accretion solutions considered in the 
present work, 
$r_h$ and $\left[u,c_s,du/dr,dc_s/dr\right]$
(evaluated on $r_h$) is determined 
using the initial boundary conditions defined by the triad
$\left[{\cal E},\lambda,\gamma\right]$ for the polytropic accretion and the diad 
$\left[T,\lambda\right]$ for the isothermal accretion, where ${\cal E},\lambda,\gamma$
and $T$ are the specific total conserved energy, specific conserved angular 
momentum, the adiabatic index ($\gamma=c_p/c_v$, where $c_p$ and $c_v$ are 
specific heats at constant pressure and volume, respectively), and the bulk
ion temperature of the accreting matter, respectively. 
Extremisation of $\kappa$, as will be shown in the 
subsequent sections, nonlinearly depends on 
$\left[\mathcal{E}, \lambda, \gamma\right]$ and on $\left[T,\lambda\right]$
for the adiabatic and the isothermal flows, respectively. One thus needs to 
explore the three dimensional parameter space spanned by 
$\left[\mathcal{E}, \lambda, \gamma\right]$ and the two dimensional parameter space spanned by 
$\left[T,\lambda\right]$
to apprehend what values of the initial boundary conditions are favoured for the 
extremisation of $\kappa$. This might help to enhance the possibility 
of obtaining the observable 
signature of the analogue radiation. One also needs 
to understand which one out of these three flow configurations 
favours the production of 
reasonably large value of $\kappa$.

This enables one to provide a `calibration space' spanned by various astrophysically 
relevant parameters governing the flow, for which the extremisation of $\kappa$ 
can be performed.
This also provides a comprehensive idea about the influence 
of 
the geometric configuration of the 
black hole accretion flow on the extremisation process of $\kappa$.

At this point, it is important to clarify that the present work does not 
make any attempt to understand the thermal properties of the Hawking like effects.
We do not intend to analyze the analogue radiation -- it's 
origin, propagation and observational manifestation. We rather concentrate 
to explore the underlying sonic geometry through the detailed study of 
the dependence of $\kappa$ on various factors 
governing the stationary background configuration. This does not 
involve the detailed analysis of quantum acoustic Hawking process at least at this stage.
We do not deal with the quantization process 
of the associated phonon field. To 
accomplish that task, one needs to demonstrate that the effective 
action for the acoustic perturbation is equivalent to a field theoretic action in 
curved space, and the associated commutation relation as well as the 
dispersion relation will directly follow \cite{unr95,us07}. 
Such considerations are rather involved and is clearly beyond the scope of 
our present work. Our main motivation is rather to employ the analogy 
to describe the classical perturbation of the fluid flow in terms of a 
field satisfying the wave equation in an effective geometry and to study the 
consequences relevant to a large-scale gravitating system.
We believe that $\kappa$ itself is 
a rather important entity to understand the flow structure as well 
as the associated sonic metric, irrespective of the existence of 
quantum Hawking like phenomena characterized by their feeble 
temperature too difficult to detect experimentally. 

The significant role 
of $\kappa$ in influencing the non 
negligible classical effects associated with the emergence of the 
stimulated Hawking effects through the modified dispersion relations
at the sonic horizon has recently been emphasized from the 
theoretical front
\cite{leonhardt-hawking-radiation-dispersive-media,hawking-radiation-in-dispersive-media-review-article}, as well 
as within the experimental framework in the laboratory set up
\cite{2008-njp,2010-njp,hydraulic-jump-white-hole-2011-pre,silke-unruh-experiment-prl}. 
The deviation of the Hawking like effects in a dispersive medium from the original 
Hawking effect is sensitive to the spatial velocity gradient corresponding to 
the stationary solutions of the background fluid flow
\cite{leonhardt-hawking-radiation-dispersive-media,hawking-radiation-in-dispersive-media-review-article}. 
For relativistic accretion onto 
a Schwarzschild black hole, the 
expression for $\kappa$ is found to be an analytical function of the 
space gradient of the steady state bulk velocity of the background fluid.
Such velocity gradient influences the universality of the Hawking like radiation (as well as the departure 
from it), and various other properties of the 
anomalous scattering of the acoustic mode due to the modified dispersion 
relation at the acoustic horizon. 

One of the main importances of our work is 
to identify a natural large-scale gravitating relativistic system where 
there is a probability to estimate the aforementioned deviation.
For such a system the {\it exact value} of the 
space gradient of the flow velocity can {\it explicitly} 
be computed in terms of realistic, observationally measurable, 
astrophysically relevant physical entities. It is surely a step 
ahead of some abstract theoretical calculation as we believe. 
Existing works which study
the anomalous dispersion relation, consider the 
gradient of the bulk velocity only and the space gradient of the 
sound speed is not taken into account in any such literature.
For adiabatic flow, the speed of 
propagation of the linear perturbation embedded within the fluid is 
a position dependent quantity, the role of the gradient of the 
sonic velocity in influencing the estimation of the deviation 
of the Hawking like effect from universality cannot be underestimated. 
In our work we calculate the space gradient of the sonic 
velocity in terms of the observationally obtainable 
physical quantities and include such factors 
in the calculation of $\kappa$. Our work can contribute to 
enrich the formalism as presented in 
\cite{leonhardt-hawking-radiation-dispersive-media,hawking-radiation-in-dispersive-media-review-article}
in a more realistic way. 

In what follows, we describe our overall scheme for the computation of $\kappa$
in terms of various flow geometries and for different thermodynamic equations of state. 

Hereafter, any relevant distance will be scaled in units of $GM_{BH}/c^2$
and any velocity will be scaled by the velocity of light in vacuum, $c$,
where $M_{BH}$ represents the mass of the black hole and $G$ represents the universal gravitational constant.

For adiabatic accretion, the equation of state of the form
\begin{equation}
p=K\rho^{\gamma} \label{anal18}
\end{equation}
is considered to describe the flow. 
$\gamma$ is assumed to be constant 
throughout the flow in the steady state. A more realistic flow
model, however, perhaps requires the implementation of a non constant
polytropic index having a functional dependence on the radial distance of
the form $\gamma\equiv\gamma (r)$ \cite{mstv2004,rc2006, mm2007, mb2011, md2012}.  
We, nevertheless, have performed our calculations for a reasonably wide spectrum of $\gamma$
and thus believe that the whole astrophysically relevant range of polytropic indices is covered in our analysis.
The proportionality constant $K$ in eq. (\ref{anal18}) is a measure of 
the specific entropy of the accreting fluid provided no additional entropy generation takes place.

Isothermal accretion is assumed to be described by the following equation of state
\begin{equation}
p=\rho{c_s^2}=\frac{\cal R}{\mu}\rho{T}=\frac{\rho{\kappa_B}T}{{\mu}m_H}
\label{anal22}
\end{equation}
${\cal R},\kappa_B,T,\mu$ and $m_H$ are the universal gas constant, the Boltzmann constant, the isothermal flow temperature,
the reduced mass and the mass of the Hydrogen atom, respectively.
$c_s$ in the above equation represents the position independent isothermal sound speed
which implies that $dc_s/dr=0$ identically. 
For isothermal accretion, only the space
gradient of the bulk advective velocity, and not that of the 
speed of propagation of the acoustic perturbation, contributes
to the estimation of the acoustic surface gravity.

For energy momentum tensor corresponding to an ideal fluid 
considered in a Boyer-Lindquist \cite{boyer} line element for 
a non rotating black hole, 
we demonstrate (see subsequent sections for detailed 
derivation) that $\kappa$ can 
be expressed as 
\begin{equation}
\kappa=\left[\frac{r-2 }{r^2 \left(1-c_s{}^2\right)}
\sqrt{r^2-\lambda^2\left(1-\frac{2}{r}\right)}
\left(\frac{du}{dr}- \frac{dc_s}{dr}\right)\right]_{r_h}
=f\left[r,u,c_s,du/dr,dc_s/dr\right]_{\rm r=r_h}
\label{anal17}
\end{equation}

In subsequent sections, we provide the expressions for 
$\left[u,c_s,du/dr,dc_s/dr\right]_{r_h}$ for three
different flow configurations for the adiabatic as well 
as the isothermal (for which $dc_s/dr$ will vanish 
everywhere, including at the acoustic horizon, for obvious reason)
equation of state. We use those values to 
study the dependence of $\kappa$ on various flow parameters as well 
as on various flow geometries. 
\section{Overall solution scheme}
\label{section3}
\noindent
We consider low angular momentum
axially symmetric accretion and the viscous transport of
angular momentum has not been taken into account.
Such low angular momentum inviscid flow is 
not a theoretical abstraction. For astrophysical systems, such 
sub-Keplerian weakly rotating flows are exhibited in
various physical situations, such as detached binary systems
fed by accretion from OB stellar winds \cite{ila-shu,liang-nolan},
semi-detached low-mass non-magnetic binaries \cite{bisikalo},
and super-massive black holes fed by accretion from slowly rotating 
central stellar clusters (\cite{ila,ho} and references therein).
Even for a standard Keplerian accretion disc, turbulence may produce 
such low angular momentum flow (see, e.g.,~\cite{igu}, and references 
therein). Reasonably large radial
advective velocity for the slowly rotating sub-Keplerian flow
implies that the infall time scale is considerably small compared to the
viscous time scale for the flow profile considered in this work.
Large radial velocities even at larger distances are due to the fact
that the angular momentum content of the accreting fluid
is relatively low \cite{belo,belo1,2003}. The assumption of
inviscid flow for the accretion profile
under consideration is thus justified from an astrophysical point of view.
Such inviscid configuration has also been addressed by other authors using
detailed numerical simulation works \cite{2003,agnes}.
The general relativistic 
Euler and the continuity equations are thus obtained 
using the vanishing of the 
four divergence of the energy momentum 
tensor of an {\it ideal fluid}. 

Considering the flow to be steady 
and the steady state to be a stable state\footnote{One can perform 
a linear stability analysis to ensure that 
the steady axially symmetric inviscid flow is stable, see
\cite{deepika-schwarzschild}.}, the time independent 
Euler and the continuity equations will 
then be integrated to obtain the respective integrals 
of motion, since the time independent Euler and the continuity equations 
are examples of first order ordinary homogeneous differential equations
in advective velocity. The integral solution of the Euler 
equation provides the conserved total specific energy 
(denoted by ${\cal E}$ in this work) as the first integral of 
motion for polytropic accretion. For isothermal flow, the 
corresponding first integral of motion (denoted by $\xi$ 
in this work) cannot be identified with the specific energy 
of the flow since energy exchange with the surrounding is required
to maintain the space invariance of the bulk temperature.
The first integral of motion obtained 
from the Euler equation does not depend on the geometric 
configuration of the flow. 

The equation of 
continuity implies the conservation of mass and hence its 
integral solution will provide the mass accretion rate 
(denoted by ${\dot {M}}$ in this work)
as another first integral of the motion. Explicit expression for the 
mass accretion rate may not depend on the equation of state 
used and is found to be a function of the flow thickness. ${\dot {M}}$
explicitly depends on the flow 
geometry. For polytropic accretion, we will 
have three different expressions for ${\dot {M}}$
for different flow geometries
and for the isothermal accretion will have same set of 
expressions for ${\dot {M}}$ for flow with constant thickness 
and conical flow, but the explicit expression will be different for 
accretion in the vertical equilibrium. This is due to the fact that 
for accretion in vertical equilibrium the expression for the flow 
thickness comes out to be a function of the corresponding sound speed. 
We solve six different cases in this work, 
three different flow models for a particular energy first integral 
for the polytropic flow as well as for the first integral corresponding 
to the isothermal flow. 

Once the first integrals of motion are obtained, we find the 
space gradient of the dynamical velocity $u$ and that of the 
sonic velocity $c_s$ (for isothermal flow $c_s$ is 
position independent) and perform the critical point analysis 
to find out the critical point(s) of the flow. For all flow models 
other than the flow in vertical equilibrium, the critical points 
coincide with the sonic points. For flow in vertical equilibrium, 
critical surfaces are not isomorphic with the sonic surfaces and the 
integral flow solutions are to be used to find the sonic point
(location of the acoustic horizon) by integrating the flow 
starting from the corresponding saddle type critical points. 
We study the dependence of $\kappa$ on 
$\left[{\cal E}, \lambda,\gamma\right]$ and on 
$\left[T,\lambda\right]$ for the polytropic as well as 
for the isothermal accretion, respectively. We compare 
such dependence for three different flow profiles.

Hereafter, the subscripts CH, CF and VE will indicate that the 
quantities are evaluated/expressions are formulated for flow 
with constant height (CH), for quasi spherical conical flow (CF),
and for flow in hydrostatic equilibrium along the vertical 
direction (VE), respectively.

\section{Configuration of the background fluid flow}
\label{section4}
\noindent
We consider a (3+1) stationary axisymmetric space-time endowed with two
commuting Killing fields, within which the dynamics of the background fluid will be 
studied. For the energy momentum tensor of any ideal fluid with certain 
equation of state, the combined equation of motion in such a configuration 
can be expressed as
\begin{equation}
v^{\mu}\nabla_{\mu}v^{\nu}+\frac{c_s^2}{\rho}\nabla_{\mu}\rho\left(g^{\mu\nu}+v^{\mu}v^{\nu}\right)=0,
\label{anal9}\end{equation}
$v^\mu$ being the velocity vector field defined on the manifold constructed by 
the family of streamlines. The normalization condition for such velocity field yields
$v^\mu{v_\mu}=-1$, and $c_s$ is the speed of propagation of the acoustic perturbation 
embedded inside the bulk flow. $\rho$ is the local rest mass energy density. The 
local timelike Killing fields $\xi^{\mu}\equiv\left(\partial{/\partial{t}}\right)^\mu$  
and $\phi^{\mu}\equiv\left(\partial{/\partial{\phi}}\right)^\mu$ are the generators of the 
stationarity (constant specific flow energy is the outcome) and axial symmetry, 
respectively. 

In general, the acoustic ergosphere and the acoustic event horizon do not co-incide.
However, for a radial flow onto a sink placed at the origin of a stationary axisymmetric
geometry they do (see, e.g., \cite{bil99,abd06} for detail discussion), 
since only the radial component of the flow velocity $u=u_\perp$ remains 
non zero everywhere. In this work we consider accretion flow with radial advective 
velocity $u$ confined on the equatorial plane. The flow will be assumed to have finite radial 
spatial velocity $u$ (the advective flow velocity as designated in usual astrophysics
literature \cite{fkr02,kato-book}) defined on the equatorial plane of the axisymmetric 
matter configuration. We focus on stationary solutions of the fluid dynamic 
equations (to determine the stationary background geometry) and hence consider only 
the spatial part of such advective velocity. Considering $v$ to be the magnitude of the three velocity, $u$
is the component of three velocity perpendicular to the set of timelike
hypersurfaces $\left\{\Sigma_v\right\}$ defined by $v^2={\rm constant}$.

The
local radial Mach number $M$ of the accreting fluid is defined as the ratio
of the radial component of the local dynamical flow velocity to that of
the propagation
of the acoustic perturbation embedded inside the accreting matter --
$M=u/c_s$.
The flow will be locally subsonic or supersonic according to $M < 1$
or $ >1$. The flow is transonic if at any moment
 it crosses the $M=1$ hypersurface. This happens when a subsonic to
supersonic or supersonic to
subsonic transition takes place either continuously or discontinuously.
Such a point where such crossing
takes place continuously is called a sonic point,
and where such transition takes place discontinuously is called a shock
or a discontinuity.
The particular value
of the radial distance $r$
for which {$M=1$}, is referred as the
transonic point or the sonic point, and will be denoted by
$r_s$ hereafter. $r_s$ and $r_h$ is thus 
identical for a transonic system. For $r<r_s$,  infalling matter becomes supersonic. 
Any
acoustic perturbation created in such a region is destined to be dragged
towards the black hole, and can not escape to the domain
$r>r_s$. In other words, any co-moving observer from
$r < r_s$ region can not communicate with any observer (co-moving or stationary)
located in the sub-domain
$r>r_s$ by sending any signal which travels with velocity
$v_{\rm signal}{\le}c_s$, where $c_s$ is defined as the velocity of
propagation of the acoustic perturbation (the sound speed) embedded in the
moving fluid. Hence the hypersurface through
$r_s$ is generated by the acoustic null geodesics,
i.e., by the phonon trajectories, and is actually an
acoustic horizon for stationary
configuration,
which is produced when accreting fluid makes a
transition from subsonic ({$M < 1$}) to the supersonic
({$M > 1$}) state.

At a distance far away from the black hole, accreting material almost
always remains subsonic (except possibly for the supersonic
stellar wind fed accretion) since it possesses negligible dynamical
flow velocity. On the other hand, the flow velocity will approach
the velocity of light $c$ while crossing the event horizon, while the maximum
possible value of sound speed, even for the steepest possible equation
of state, would be $c/\sqrt{3}$ \cite{fkr02,kato-book}, resulting $M>1$ close to the
event horizon.
In order to satisfy such inner boundary condition imposed by the
event horizon, accretion onto black holes
exhibit transonic properties in general \cite{liang-thomson}.

For a transonic flow as perceived within the aforementioned configuration, the collection of the 
sonic points (where the radial Mach number, the ratio of the advective velocity and the 
speed of propagation of the acoustic perturbation in the radial direction, becomes
unity) at a specified radial distance forms the acoustic horizon, the generators of which 
are the phonon trajectories. An axially symmetric transonic black hole accretion can thus 
be considered as a natural example of the classical analogue gravity model which contains 
two different horizons, the gravitational (corresponding to the accreting black hole) as 
well as the acoustic (corresponding to the transonic fluid flow). 

To describe the flow structure in further detail, the energy momentum tensor 
of an ideal fluid of the form
\begin{equation}
T^{\mu\nu}= (\epsilon+p)v^{\mu}v^{\nu}+pg^{\mu\nu}
\label{anal10}
\end{equation}
is considered in a Boyer-Lindquist \cite{boyer} line element normalized for $G=c=M_{BH}=1$ and 
$\theta=\pi/2$ as defined below \cite{nt73}
\begin{equation}
ds^2=g_{{\mu}{\nu}}dx^{\mu}dx^{\nu}=-\frac{r^2{\Delta}}{A}dt^2
+\frac{A}{r^2}\left(d\phi-\omega{dt}\right)^2
+\frac{r^2}{\Delta}dr^2+dz^2\,,
\label{anal11}
\end{equation}
where
\begin{equation}
\Delta=r^2-2r+a^2, A=r^4+r^2a^2+2ra^2,\omega=2ar/A\,,
\label{anal12}
\end{equation}
$a$ being the Kerr parameter related to the black holes spin angular momentum.
The required metric elements are:
\begin{equation}
g_{rr}=\frac{r^2}{\Delta},~g_{tt}=
\left(\frac{A\omega^2}{r^2}-\frac{r^2{\Delta}}{A}\right),~
g_{\phi\phi}=\frac{A}{r^2},~ g_{t\phi}=g_{\phi{t}}=-\frac{A\omega}{r^2}\,.
\label{anal13}
\end{equation}
The specific angular
momentum $\lambda$ (angular momentum per unit mass)
and the angular velocity $\Omega$ can thus be expressed as
\begin{equation}
\lambda=-\frac{v_\phi}{v_t}, \;\;\;\;\;
\Omega=\frac{v^\phi}{v^t}
=-\frac{g_{t\phi}+\lambda{g}_{tt}}{{g_{\phi{\phi}}+\lambda{g}_{t{\phi}}}}\, .
\label{anal14}
\end{equation}
We also define
\begin{equation}
B=g_{\phi\phi}+2\lambda{g_{t\phi}}+\lambda^2{g_{tt}}\,,
\label{anal15}
\end{equation}
which will be used in the subsequent sections to calculate the
value of the acoustic surface gravity.

For flow onto a Schwarzschild black hole, one can obtain the 
respective metric elements (and hence, the expression for $\lambda$ and $B$ 
thereof) by substituting $a=0$ in eq. (\ref{anal12} - \ref{anal15}). 
We construct a Killing vector $\chi^\mu = \xi^\mu+\Omega{\phi^\mu}$
where the Killing vectors $\xi^\mu$ and $\phi^\mu$ are the two
generators of the temporal and axial 
isometry groups, respectively. Once $\Omega$ is computed at the acoustic horizon
$r_h$, 
$\chi^\mu$ becomes null on the transonic surface. The norm of the
Killing vector $\chi_\mu$ may be
computed as
\begin{equation}
\sqrt{\left|\chi^\mu{\chi_\mu}\right|}
=\sqrt{\left(g_{tt}+2\Omega{g_{t\phi}}+\Omega^2{g_{\phi\phi}}\right)}
=\frac{\sqrt{{\Delta}B}}{g_{\phi{\phi}}+{\lambda}g_{t{\phi}}}\,.
\label{anal16}
\end{equation}
Hence the explicit form of the acoustic surface gravity for relativistic flow 
onto a Schwarzschild black hole looks like
\begin{equation}
\kappa=\left[\frac{\sqrt{r^2-2 r}}{r^2 \left(1-c_s{}^2\right)}\sqrt{\frac{g_{\phi \phi }+
\lambda ^2 g_{\text{tt}}}{g_{\text{rr}}}}\left(\frac{du}{dr}-
\frac{dc_s}{dr}\right)\right]_{r_h}
\label{anal17}
\end{equation}

\section{The first integrals of motion}
\label{section5}
\noindent
Vanishing of the four divergence of the energy 
momentum tensor provides the general relativistic version of the Euler equation
\begin{equation}
T^{\mu\nu}_{;\nu}=0.\label{anal23}
\end{equation}
whereas the corresponding continuity equation is obtained from
\begin{equation}
\left(\rho v^{\mu}\right)_{;\nu}=0.\label{anal24}
\end{equation}

The time independent part of the linear momentum conservation equation
(Euler equation) is a 
first order homogeneous differential equation. Its integral solution will provide a constant 
of motion (first integral of motion) for whatever equation of state is used to describe the 
accreting matter. Such first integral of motion, however, cannot formally be identified with the 
total energy of the background fluid flow for any equation of state other than the polytropic 
one. 

\subsection{Integral solution of the linear momentum conservation equation}
\label{label5.1}
\subsubsection{Polytropic accretion}
\label{label5.1.1}
\noindent
For polytropic accretion, the specific enthalpy $h$ is formulated as \begin{equation}
h=\frac{p+\epsilon}{\rho},\label{anal19}
\end{equation}
where the energy density $\epsilon$ includes the rest mass density 
and internal energy and is defined as \begin{equation}
\epsilon=\rho +\frac{p}{\gamma-1}
\label{anal20}
\end{equation}
The adiabatic sound speed $c_s$ is defined as \begin{equation}
c_s^2=\left(\frac{\partial p}{\partial \epsilon}\right)_{\rm constant~enthalpy}
\label{anal21}
\end{equation}
At constant entropy, the enthalpy can be expressed as
\begin{equation}
{h} = \frac{\partial \epsilon}{\partial {\rho}}
\label{anal21a}
\end{equation}
and hence 
\begin{equation}
h = \frac{\gamma - 1}{\gamma - \left(1 +c_{s}^2\right)}
\label{anal21b}
\end{equation}

Contracting eq. (\ref{anal23}) with $\phi^{\mu}$ one obtains (since $\phi^\nu{p}_{,\nu}=0,
\phi^\mu={\delta}^\mu_\phi$, and $g_{\mu{\lambda};\nu}=0$)
\begin{equation}
\left[\phi_{\mu}hv^{\nu}\right]_{;\nu}=0. \label{anal25}
\end{equation}
Since $\phi_{\mu}hv^{\mu}=hv_{\phi}$, the angular momentum per baryon for the
axisymmetry flow is conserved. 
Contraction of eq. (\ref{anal23}) with $\xi^{\mu}$ provides
\begin{equation}
\xi^{\mu} \left[\xi_{\mu}T^{\mu\nu}_{;\nu}=0\right]
\label{anal26}
\end{equation}
from where the quantity $hv_t$ comes out as one of the first integrals of motion of the 
system. $hv_t$ is actually the relativistic version of the 
Bernouli's constant \cite{anderson}
and can be identified with the total specific energy of the general relativistic 
ideal fluid ${\cal E}$ (see, e.g., \cite{das-czerny-2012-new-astronomy} and references therein) scaled 
in units of the rest mass energy.
Hence ${\cal E}=hv_t$

From the normalization condition $v^{\mu}v_{\mu}=-1$ one obtains 
\begin{equation}
v_t=\sqrt{\frac{g_{t\phi}^2-g_{tt}g_{\phi\phi}}{(1-\lambda\Omega)(1-u^2)(g_{\phi\phi}+\lambda g_{t\phi})}}
\label{anal29}
\end{equation}

We thus obtain
\begin{equation}
{\cal E} = \frac{\gamma - 1}{\gamma - \left(1 +c_{s}^2\right)}
           \sqrt{\frac{g_{t\phi}^2-g_{tt}g_{\phi\phi}}{(1-\lambda\Omega)(1-u^2)(g_{\phi\phi}+\lambda g_{t\phi})}}
\label{anal30}
\end{equation}
The exact form for ${\cal E}$ will depend on the space time structure 
appearing in the expression for ${\cal E}$ through the metric elements. 
It will not depend on the matter geometry since the accretion is assumed to be non self gravitating.
For isothermal flow, the total specific flow energy does not remain 
constant, rather the first integral of motion obtained by integrating the 
relativistic Euler equation has a different algebraic form which can not be identified with the 
total energy of the system. 

\subsubsection{Isothermal flow} 
\label{section5.1.2}
\noindent
For isothermal flow, the system has to dissipate energy to keep the temperature constant. 
The isotropic pressure is proportional to the energy density through
\begin{equation}
p = c_s^2 {\epsilon}
\label{anal31}
\end{equation}
From the time part of eq. (\ref{anal23}), one obtains 
\begin{equation}
\frac{\d v_t}{v_t}=-\frac{\d p}{p+\epsilon}
\label{anal32}
\end{equation}
Using the definition of enthalpy, the above equation may be re-written as 
\begin{equation}
\frac{\d v_t}{v_t}=-\frac{1}{h}\frac{\d p}{\d \rho}\frac{\d \rho}{\rho}
\label{anal33}
\end{equation}
Since the isothermal sound speed can be defined as (see, e.g., \cite{ydl96} and 
references therein) 
\begin{equation}
c_s=\sqrt{\frac{1}{h}\frac{\d p}{\d \rho}}
\label{anal34}
\end{equation}
we obtain 
\begin{equation}
\ln v_t=-c_s^2\ln \rho + A, \textrm{ where $A$ is a constant}
\label{anal35}
\end{equation}
Which further implies that 
\begin{equation} 
{v_t}{\rho^{c_s^2}} = \xi
\label{anal36}
\end{equation}
Hence  $\xi$ is the first integral of motion for the isothermal 
flow, which is not to be confused with the total conserved specific energy 
${\cal E}$. 

Owing to the Clayperon-Mendeleev  equation \cite{bazarov,gibbs}
\begin{equation}
c_s=\sqrt{\frac{k_B}{\mu m_H} T}
\label{anal74}
\end{equation}
space invariance of temperature requires the sound speed
to be position independent for isothermal accretion,
and hence the space gradient of the speed of propagation
of the isothermal perturbation does not contribute to the estimation of the
acoustic surface gravity. The expression for the first integral of motion obtained from the
integral solution of the Euler equation is independent of the geometrical configuration of
matter as already discussed, and is found to be
\begin{equation}
\xi=\frac{r^2(r-2)}{(r^3-(r-2) \lambda ^2) (1-u^2)} \rho^{2c_s^2}
\label{anal73}
\end{equation}

\subsection{Integral solution of the mass conservation equation}
\label{section5.2}
\noindent
For $g\equiv \det(g_{\mu\nu})$, the mass conservation equation (\ref{anal24}) implies
\begin{equation}
\frac{1}{\sqrt{-g}}(\sqrt{-g}\rho v^{\mu})_{,\mu}=0,\label{anal37}
\end{equation}
which further leads to 
\begin{equation}
\left[(\sqrt{-g}\rho v^\mu)_{,\mu}d^4 x=0\right]
\label{anal38}
\end{equation}
$\sqrt{-g}d^4 x$ being the co-variant volume element. We assume 
that there is no convection current along any non equatorial direction,
and hence no non-zero terms involving $v^{\theta}$ (for spherical polar co-ordinate) or
$v^z$ (for flow studied within the framework of cylindrical co-ordinate) should 
become significant. This assumption leads to the condition
\begin{equation}
\partial_r(\sqrt{-g}\rho v^r)\d r \d\theta\d\phi = 0,\label{anal39}
\end{equation}
for the stationary background flow studied using the spherical polar co-ordinate 
$\left(r,\theta,\phi\right)$ and 
\begin{equation}
\partial_r(\sqrt{-g}\rho v^r)\d r \d z\d\phi = 0,\label{anal40}
\end{equation} 
for such flow studied using the cylindrical co-ordinate $\left(r,\phi,z\right)$.

We integrate eq. (\ref{anal39}) for $\phi=0\rightarrow 2\pi$ and 
$\theta=-H_{\theta}\rightarrow H_{\theta}$; $\pm H_{\theta}$ being
the value of the polar co-ordinates above  and below the equatorial 
plane, respectively, for a local flow half thickness $H$, to 
obtain the conserved mass accretion 
rate $\dot{M}$ in the equatorial plane. The integral solution of the 
mass conservation equation -- the mass accretion rate 
$\dot{M}$ -- comes out to be another first integral of 
motion for our stationary background fluid configuration. For 
conical wedge shaped flow studied in the spherical polar co 
ordinate, $2H/r$ remains constant. Flow with such geometric 
configuration was first studied by \cite{az1981} and followed 
by \cite{blaes87} for 
pseudo-Schwarzschild flow geometry under the influence of 
the Paczy\'nski \& Wiita \cite{pw80} pseudo-Schwarzschild
Newtonian like black hole potential. The relativistic version 
for such flow has further been studied by 
\cite{lu85,lu86,lyy95,gammie-popham,popham-gammie,lyyy97a,lyyy97b,ydl96,ly98,lugu}.

In a similar fashion, eq. (\ref{anal40}) can be 
integrated for $z=-H_z\rightarrow H_z$ (where $\pm H_z$ is the
local half thickness of the flow) symmetrically over and below 
the equatorial plane for axisymmetric accretion studied using the
cylindrical polar co-ordinate to obtain the corresponding 
mass accretion rate on the equatorial plane. Contrary to the 
first integral of motion obtained by integrating the Euler equation 
for a particular thermodynamic equation of state,    
the expression for the mass 
accretion rate does not explicitly depend on the equation 
of state, but is different for different geometric configuration of the 
matter distribution. The general expression for the mass accretion rate 
can be provided as
\begin{equation}
\dot{M} = \rho v^r\mathcal{A}(r)
\label{anal41}
\end{equation}
$\mathcal{A}(r)$ being the two dimensional surface area 
having surface topology ${\mathbb{R}^1}\times{\mathbb{R}}^1$ 
or ${\mathbb{S}^1}\times{\mathbb{S}}^1$ through which the 
inward mass flux is estimated in the steady state. For 
${\mathbb{S}^1}\times{\mathbb{S}}^1$ (and for not so large 
value of $\theta$), $\mathcal{A}(r)=4\pi H_{\theta}r^2$, 
and for ${\mathbb{R}^1}\times{\mathbb{R}}^1$ (axisymmetric
accretion studied using the cylindrical co ordinate), 
$\mathcal{A}(r)=4\pi H_z r$.

In subsequent sections, we provide the explicit expressions for the 
conserved mass accretion rate and related quantities 
for three different flow geometries
\section{Stationary transonic accretion solutions}
\label{section6}
\subsection{polytropic accretion}
\label{section6.1}
\noindent
We substitute the values of the corresponding metric elements 
and of $\Omega$ in 
eq. ({\ref{anal30}) and obtain
\begin{equation}
{\cal E} = -\frac{\gamma -1}
{(\gamma -(1+c_s{}^2))}\sqrt{\frac{(1-\frac{2}{r})}{(1-\frac{\lambda ^2}{r^2}(1-\frac{2}{r}))(1-u^2)}}
\label{anal42}
\end{equation}
In what follows, we shall illustrate the procedure to obtain the stationary transonic flow 
solutions for flow with constant thickness. We shall then provide the corresponding similar expressions for 
other flow geometries. 
\subsubsection{Flow with constant thickness}
\noindent
As stated in the paragraphs preceding eq. (\ref{anal41}),
we integrate the continuity equation to obtain the 
conserved mass accretion rate to be 
\begin{equation}
\dot{M}_{\rm CH}=2 \pi \rho \frac{u\sqrt{1-\frac{2}{r}}}{\sqrt{1-u^2}} r H
\label{anal43}
\end{equation}
$H$ being the constant disc height.

Equations (\ref{anal42} -- \ref{anal43}) can not directly be solved simultaneously since it contains three unknown variables
$u,c_s$ and $\rho$, all of which are functions of the radial distance $r$. Any accretion variable 
from the triad $\left[u,c_s,\rho\right]$ has to 
be eliminated in terms of the other two. We are, however, interested to study the radial Mach number profile to identify the
location of the acoustic horizon (the radial distance at which $M$ becomes unity), and hence the study of the radial 
variation of $u$ and $c_s$ are of prime interest in this case. We would thus like to express $\rho$ in
terms of $c_s$ and other related constant quantities. To accomplish the aforementioned task, we make a
transformation ${\dot \Xi}={\dot M}K^{\frac{1}{\gamma-1}}{\gamma^{\frac{1}{\gamma-1}}}$. Employing the
definition of the sound speed $c_s^2=\left(\frac{\partial{p}}{\partial{\epsilon}}\right)_{\rm Constant~Entropy}$
as well as the equation of state used to describe the flow, the expression for
${\dot \Xi}$ can further be elaborated as
\begin{equation}
\dot{\Xi}_{\rm CH}=2{\pi} \frac{u\sqrt{1-\frac{2}{r}}}{\sqrt{1-u^2}} rc_s^{\frac{2}{\gamma -1}}(\frac{\gamma -1}{\gamma - (1+c_s^2)})^\frac{1}{\gamma -1}H
\label{anal44}
\end{equation}
The entropy per particle $\sigma$ is related to $K$ and $\gamma$ as
\cite{landau-lifshitz-stat-mech-book}
$$
\sigma=\frac{1}{\gamma -1}\log K+
\frac{\gamma}{\gamma-1}+{\rm constant}
\label{anal45}
$$
where the constant depends on the chemical composition of the
accreting material.
The above equation implies that $K$
is  a measure of the specific entropy of the accreting matter.
We thus interpret ${\dot {\Xi}}$ as the measure of the total
inward entropy flux associated with the accreting material and label
${\dot {\Xi}}$ to be the stationary entropy accretion rate.
The concept of the entropy accretion rate was first introduced in
\cite{az1981,blaes87} to obtain the stationary transonic solutions
of the low angular momentum non relativistic axisymmetric accretion
under the influence of the Paczy\'nski and Wiita \cite{pw80}
pseudo-Schwarzschild potential onto a non rotating black hole.

The conservation equations for ${\cal E}, {\dot M}$ and
${\dot \Xi}$ may simultaneously be solved
to obtain the complete accretion profile on the 
radial Mach number vs radial distance phase space, 
see, e.g., \cite{das07,das-czerny-2012-new-astronomy}
for the depiction of several such phase portraits. 

The relationship between the space gradient of the acoustic velocity
and that of the advective velocity can now be established by differentiating
eq. (\ref{anal44})
\begin{equation}
\left[\frac{dc_s}{dr}\right]_{\rm CH}= -\frac{\gamma -1}{2}\frac{\left\{\frac{1}{u}+
\frac{u}{1-u^2}\right\}\frac{\text{du}}{\text{dr}}+\left\{\frac{1}{r}+
\frac{1}{r^2(1-\frac{2}{r})}\right\}}{\frac{1}{c_s}+\frac{c_s}{\gamma -(1+c_s^2)}}
\label{anal46}
\end{equation}

Differentiation of eq. (\ref{anal42}) with respect to the
radial distance $r$ provides another relation between
$dc_s/dr$ and $du/dr$. We substitute $dc_s/dr$ as obtained from
eq. (\ref{anal46}) into that relation and finally obtain the expression for the
space gradient of the advective velocity as
\begin{equation}
\left[\frac{du}{dr}\right]_{\rm CH} = 
\frac{c_s{}^2\left\{\frac{1}{r}+\frac{1}{r^2(1-\frac{2}{r})}\right\}-f_2(r,\lambda )}{(1-c_s{}^2)\frac{u}{1-u^2}-\frac{c_s{}^2}{u}} 
= \frac{\it N_1}{\it D_1}
\label{anal47}
\end{equation}
where 
\begin{subequations}
\begin{align}
f_2(r,\lambda) = -\frac{\lambda ^2}{r^3}\left\{\frac{1-\frac{3}{r}}{1-\frac{\lambda ^2}{r^2}(1-\frac{2}{r})}\right\}+\frac{1}{r^2(1-\frac{2}{r})} \label{anal48b} \\
\nonumber \\
\text{We define another quantity $f_1(r,\lambda)$ which shall be used later,} \nonumber \\
f_1(r,\lambda)=\frac{3}{r}+\frac{\lambda ^2}{r^3}\left\{\frac{1-\frac{3}{r}}{1-\frac{\lambda ^2}{r^2}(1-\frac{2}{r})}\right\}
\label{anal48a}\\
\nonumber
\end{align}
\end{subequations}
Eq. (\ref{anal46} -- \ref{anal47}) can now be identified with a
set of non-linear first order differential equations representing
autonomous dynamical systems \cite{gkrd07}, and their integral solutions provide
phase trajectories on the radial Mach number $M$ vs the radial distance $r$
plane. The `regular' critical point conditions for these integral solutions are
obtained by simultaneously making the numerator and the denominator of
eq. (\ref{anal47}) vanish. The aforementioned critical point
conditions may thus be expressed as
\begin{equation}
\left[u=c_s\right]_{r_c}, ~~~ \left[c_s\right]_{r_c}=
\sqrt{\frac{f_2(r_c,\lambda )}{\frac{2}{r_c}+\frac{1}{r_c{}^2(1-\frac{2}{r_c})}}}
\label{anal49}
\end{equation}
Since in this work we deal with a transonic fluid in real space for which 
the flow is continuous along the entire real line, only
the `regular' or 'smooth' critical point is 
considered, for
which $u,c_s$ as well as their space derivatives remain regular
and do not diverge. Such a critical point may be of saddle type
allowing a transonic solution to pass through it, or may be
of centre type through which no physical transonic solution can
be constructed. Other categories of critical point
include a `singular' one for which $u,c_s$ are continuous
but their derivatives diverge. All such classifications
have been discussed in detail and the criteria for a critical point to
qualify as a `regular' one which is associated with a physical
acoustic horizon has been found out in \cite{abd06}.

Equation (\ref{anal49}) provides the critical point
condition but not the location of the critical point(s). It is
necessary to solve eq. (\ref{anal42}) under the critical point condition for
a set of initial boundary conditions as defined by
$\left[{\cal E},\lambda,\gamma \right]$. The value of $c_s$ and $u$, as
obtained from eq. (\ref{anal49}), may be substituted in eq. (\ref{anal42}) to
obtain the following 11$^{th}$ degree algebraic polynomial equation for
$r=r_c$, $r_c$ being the location of the critical point
\begin{equation}
a_0 + a_1 r_c + a_2 r_c^2 + a_3 r_c^3 + a_4 r_c^4 + a_5 r_c^5 + a_6 r_c^6 + a_7 r_c^7 + a_8 r_c^8 + a_9 r_c^9 + a_{10} r_c^{10} + a_{11} r_c^{11}=0
\label{anal50}
\end{equation}
where, the coefficients $a_i$ are functions of $\left[{\cal E},\lambda,\gamma\right]$. 
The explicit form of such co-efficients are derived, and the results are presented 
in the appendices. 

A particular
set of values of $\left[{\cal E},\lambda,\gamma \right]$ will then
provide the numerical solution for the algebraic expression to obtain
the exact value of $r_c$. Astrophysically relevant domain for such 
initial boundary conditions are \cite{das-czerny-2012-new-astronomy}
defined by 
\begin{equation}
\left[1{\lsim}{\cal E}{\lsim}2,0<\lambda{\le}4,4/3{\le}\gamma{\le}5/3\right]
\label{anal52}
\end{equation}
For accretion with constant height, the 
critical point condition reveals that the advective velocity and the sound velocity 
are same at the critical point. Hence the critical surface at $r_c$ and the 
acoustic horizon $r_h$ coincide for this flow geometry. 

For an astrophysically relevant set of $\left[{\cal E},\lambda,\gamma\right]$ the critical 
point(s) of the phase trajectory can be identified, and a linearisation
study in the neighbourhood of these critical points(s) may be performed \cite{gkrd07}
to develop a classification scheme to identify the nature of the critical point(s).
Since viscous transport of angular momentum has not been taken into account in the 
present work, such critical points are either of saddle type through which a 
stationary transonic flow solution can be constructed, or of a
centre type which does not allow any transonic solution on phase portrait to 
pass through it. A complete understanding of the background stationary transonic 
flow topologies on the phase portrait will require a numerical integration of 
the non analytically solvable non linearly coupled differential equations 
describing the space gradient of the advective velocity as well as that of the 
speed of propagation of the acoustic perturbation embedded within the 
stationary axisymmetric background spacetime. 

For a particular set of $\left[{\cal E},\lambda,\gamma\right]$, solution of 
eq. (\ref{anal50}) provides either no real positive root lying outside 
the gravitational black hole horizon implying that no acoustic horizon 
forms outside the black hole event horizon (non availability of the 
transonic solution) for that value of $\left[{\cal E},\lambda,\gamma\right]$,
or provides one, two or three (at most) real positive roots lying outside 
the black hole event horizon. Typically, if only one root is found, the critical point 
is of saddle type and a mono-transonic flow profile is obtained with a single acoustic horizon 
for obvious reason. Solutions containing two critical saddle points 
imply the presence of a homoclinic orbit\footnote{A homoclinic orbit or a 
homoclinic connection is a 
bi-asymptotic trajectory converging to a saddle like orbit as time 
goes to positive or negative infinity. For our stationary 
systems, a homoclinic orbit on a phase portrait is realized as an
integral solution that re-connects a saddle type critical point to itself
and embarrasses the corresponding centre type critical point. For a detailed
description of such phase trajectories from a dynamical systems point of view,
see, e.g., \cite{js99,diff-eqn-book,strogatz}.} on the phase plot and hence such 
solutions are excluded. 

Stationary configuration with three critical points requires a somewhat detailed 
understanding. Although a full description is available in 
\cite{das-czerny-2012-new-astronomy}, we provide
a brief account over here for the sake of completeness. 
One out of the aforementioned three critical points is of 
centre type which is  circumscribed by two saddle type critical points,
see, e.g., \cite{js99} for details about the characteristic features of the 
saddle and the center type critical points}. 
With reference to the gravitational horizon, one of these saddle points 
forms sufficiently close to it, even closer than the innermost circular 
stable orbit. i.e., ISCO (see, e.g. \cite{fkr02} and \cite{kato-book}
for details about ISCO) in general, and is termed as the inner type critical
point. 
The other saddle type point, termed as the outer saddle type critical point, 
is usually formed at a fairly large distance away from the gravitational horizon. The 
inner critical point thus forms in a region of substantially strong gravitational field 
whereas the outer type critical point, in many cases, is formed in a region of 
asymptotically flat spacetime. This is because,
depending on the choice of $\left[{\cal E},\lambda,\gamma\right]$,
such a critical point can be located at a distance $10^6GM_{BH}/c^2$
(or even more) away from the gravitational horizon. 
The centre type critical point, termed as the middle critical point because of 
the fact that $r_{c}^{\rm inner}<r_c^{\rm middle}<r_c^{\rm outer}$, forms usually 
at a length scale ranging from 10 to 10$^{\rm 3-4}$ in units of $GM_{BH}/c^2$, 
depending on the value of $\left[{\cal E},\lambda,\gamma\right]$ used.
$\left[{\cal E},\lambda,\gamma\right]_{\rm mc} \subset \left[{\cal E},\lambda,\gamma\right]$
thus provides the multi-critical behaviour of stationary transonic solution. 
The parameter space spanned by $\left[{\cal E},\lambda,\gamma\right]_{\rm mc}$ can further be classified 
into two different subspaces for which the representative phase portraits are topologically 
different. Such subspaces are characterized by the relative values of the stationary 
entropy accretion rate $\dot {\Xi}$ evaluated at the inner and the outer critical points,
respectively. For ${\dot {\Xi}}_{r_c^{\rm inner}}>{\dot {\Xi}}_{r_c^{\rm outer}}$, 
accretion can have three allowed critical points, and a homoclinic orbit is generated through the 
inner saddle type critical point, whereas for ${\dot {\Xi}}_{r_c^{\rm inner}}<{\dot {\Xi}}_{r_c^{\rm outer}}$
transonic accretion can have only one saddle type (inner) critical point and the homoclinic 
orbit forms through the outer saddle type critical point. Hence {\it only} 
$\left[{\cal E},\lambda,\gamma\right]_{\rm mc}^{{\dot {\Xi}}_{r_c^{\rm inner}}>{\dot {\Xi}}_{r_c^{\rm outer}}}\subset\left[{\cal E},\lambda,\gamma\right]_{\rm mc}$ provides the multi-critical accretion configuration for which two saddle type 
and one center type (delimited between two such saddle types) critical points are available. As already mentioned, a physically 
acceptable transonic solution for inviscid accretion cannot be constructed through a centre type critical point.
A multi-critical flow with three critical points is thus a theoretical abstraction.

On the other hand, a bi-transonic accretion 
is a practically realizable configuration where the stationary transonic accretion solution passes through one inner 
and one outer saddle type sonic points. For flow geometries providing the isomorphism between the critical and the 
sonic points, such sonic points define the acoustic horizons. For flow configuration which does not 
allow such isomorphism, a sonic point can be identified on the integral stationary flow solutions corresponding to 
every saddle type critical point. For a bi-transonic solution, however, it should indeed be realized that a smooth 
stationary solution can not encounter more than one regular sonic point since once it crosses the outer type 
sonic point (for accretion) it becomes supersonic and only a subsonic solution can have access to pass through 
the inner sonic point. No continuous transonic solution can accommodate more than one acoustic horizons. 
Multi transonicity
could only be realized as a specific flow configuration where the
combination of two different otherwise smooth solutions
passing through two different saddle type critical (and hence sonic) points
are connected to each other through a discontinuous shock transition.
Such a shock has to be stationary and will be located between two sonic
points. For certain 
$\left[{\cal E},\lambda,\gamma\right]_{\rm nss}\subset\left[{\cal E},\lambda,\gamma\right]_{\rm mc}^{{\dot {\Xi}}_{r_c^{\rm inner}}>{\dot {\Xi}}_{r_c^{\rm outer}}}$
where `nss' stands for no shock solution, three
critical points (two saddle embracing a centre one)
are routinely obtained but no stationary shock
forms for the stationary transonic accretion. Hence no multi transonicity is observed even if the flow is
multi-critical, and real physical accretion solution can have access
to only one saddle type critical point (the outer one) out of the two.
Thus multi
critical accretion and multi transonic accretion are not topologically
isomorphic in general. A true multi-transonic flow can only be
realized for
$\left[{\cal E},\lambda,\gamma\right]_{\rm ss}\subset\left[{\cal E},\lambda,\gamma\right]_{\rm mc}^{{\dot {\Xi}}_{r_c^{\rm inner}}>{\dot {\Xi}}_{r_c^{\rm outer}}}$
where
`ss' stands for `shock solution', if the criteria for the energy preserving relativistic Rankine-Hugoniot 
shock \cite{rankine,hugoniot1,hugoniot2,landau,salas} for the adiabatic accretion and temperature 
preserving relativistic shock \cite{yk95,ydl96} for the isothermal accretion are met. 
In this work, however, we will not be interested to deal with the shock solutions and would 
mainly concentrate on the mono-transonic flow to study the accretion model dependence of 
the acoustic surface gravity $\kappa$. Further details will be provided in subsequent paragraphs where 
we describe the methodology of constructing the Mach number vs radial distance (measured 
from the gravitational horizon in units of $GM_{BH}/c^2$) phase portrait, see, e.g., 
figure 2 of \cite{das-czerny-2012-new-astronomy} and related discussions may be found therein for details of 
such multi-transonic shocked accretion flow configurations.

Space gradient of the advective velocity 
at the critical points for various flow
configurations can be obtained by evaluating the limiting values of $\left(du/dr\right)$
at such points using the l'H\^{o}pital's rule \cite{original-paper}. The
expressions for the critical values of $\left(du/dr\right)$ and
$\left(dc_s/dr\right)$ have been provided in the appendix.

The critical acoustic velocity gradient
$\left(dc_s/dr\right)_{\rm r=r_c}$ can also be computed by
substituting the value of $\left(\frac{du}{dr}\right)_{\rm r=r_c}$
in eq. (\ref{anal46}) and by
evaluating other quantities in eq. (\ref{anal46}) at $r_c$.
Both $\left(dc_s/dr\right)_{\rm r=r_c}$ and 
$\left(\frac{du}{dr}\right)_{\rm r=r_c}$ can 
be reduced to an algebraic expression in $r_c$ with
real coefficients that are complicated functions of $\left[{\cal E},\lambda,\gamma\right]$. 
Once $r_c$ is known for a set of values of $\left[{\cal E},\lambda,\gamma\right]$,
the critical slope for the advective velocity, i.e., the space gradient for $u$ at
$r_c$ can be computed as a pure
number, which may either be a real number providing a saddle type 
point (for stationary transonic accretion solution to
exist) or an imaginary number for providing a centre type point (no transonic solution can be found).

To obtain the Mach number vs radial distance phase plot for the stationary 
transonic accretion flow, one needs to simultaneously integrate the set of 
coupled differential equations (\ref{anal46} -- \ref{anal47}) for a 
specific set of initial boundary conditions determined by $\left[{\cal E},\lambda,\gamma\right]$.
The initial value of the space gradient of the advective velocity,
i.e., the critical velocity gradient evaluated at the critical point and 
provided in  eq. (\ref{apen11}) and the critical space gradient of the 
sound speed can be numerically iterated using the fourth order Runge - Kutta 
method \cite{numerical-recipes} to obtain the integral solutions for the 
mono-transonic as well as for the multi-transonic flow. Details of such 
numerical integration scheme, along with the representative phase plots are 
available in \cite{das-czerny-2012-new-astronomy,das07,pmdc12}. Acoustic surface gravity
is not relevant for a centre type critical point since no stationary 
transonic solution can be constructed through such points.
For flow with constant thickness and for conical flow --
both for the adiabatic as well as the isothermal accretion -- 
saddle type critical points and the sonic points are 
isomorphic.  
Since the critical surface and the acoustic horizon is identical, 
numerical construction of the integral stationary solution
is not required to calculate the corresponding acoustic surface gravity for these flow 
geometries, and 
value of $\left[u,c_s,du/dr,dc_s/dr\right]_{\rm r_c}$ is all we need
to calculate the value of $\kappa$ for the respective flow configuration. 
In such cases, for axisymmetric flow with constant height for example,
the surface gravity can be computed as
\begin{equation}
\kappa_{\rm Adiabatic}^{\rm Constant~Height}=\left| \frac{r-2}{r^2 \left(1-c_s{}^2\right)}\sqrt{r^2-\lambda ^2 \left(1-\frac{2}{r}\right)} \left[\eta_1\frac{\text{du}}{\text{dr}}+\sigma_1\right] \right|_{\rm r_c}
\label{anal56a}
\end{equation}
where
\begin{equation}
\eta_1 = {1+\frac{\gamma -\left(1+u^2\right)}{2\left(1-u^2\right)}}, ~\sigma_1={\left(\frac{\gamma -1}{2}\right)\left(\frac{\frac{1}{r}+\frac{1}{r^2\left(1-\frac{2}{r}\right)}}{\frac{1}{u}+\frac{u}{\gamma -\left(1+u^2\right)}}\right)}
\label{anal56b}
\end{equation}

For accretion in hydrostatic equilibrium along the 
vertical direction, critical points and the sonic points 
are not isomorphic. As a result, the acoustic horizon does not form on the 
critical surface. 
The location 
of the sonic point
will always be located at a radial distance 
$r_{\rm sonic} < r_{\rm critical}$ (hereafter we will designate a sonic point as $r_s$ 
instead of $r_{\rm sonic}$). Such $r_s$ is to be found out by integrating 
the expression of $\left(du/dr\right)$ and $\left(dc_s/dr\right)$ and by locating the radial 
co ordinate on the equatorial plane for which the Mach number becomes exactly equal to unity. 
Since the condition $u^2-c_s^2=0$ is satisfied at $r_s$ and not at $r_c$,
$\left[u,c_s,du/dr,dc_s/dr\right]_{\rm r_s}$ is to be used to calculate the 
corresponding value of the acoustic surface gravity 
instead of $\left[u,c_s,du/dr,dc_s/dr\right]_{\rm r_c}$ for those particular flow geometries.

It is relevant to note that the absolute value of the (constant) disc 
thickness $H$ does not enter anywhere in the expression of the acoustic surface gravity
(and hence, into the calculation of the Hawking like temperature). 
Similar result is to be obtained 
for the conical flow where the geometrical factor (the solid angle) 
representing the angular opening of the conical flow does not show 
up in the expression for $\kappa$ or $T_{AH}$ as well. This implies that 
it is only the geometrical configuration of the matter (non self gravitating) and not 
the absolute measure of the flow thickness (for constant height flow) or the ratio 
of the local height to the local radial distance (for conical flow) 
which influences the computation of the acoustic surface gravity. 
This may not be the situation where the radius dependent 
flow thickness itself is found to be 
a function of the speed of propagation of acoustic perturbation. 
\subsubsection{Conical flow model}
\label{section6.1.2}
\noindent
The mass accretion rate is calculated as 
\begin{equation}
{\dot{M}}_{\rm CF}=\Lambda \rho \frac{u\sqrt{1-\frac{2}{r}}}{\sqrt{1-u^2}} r^2
\label{anal57}
\end{equation}
$\Lambda$ is the geometric factor determining the exact shape of the flow, over which
integration of the continuity equation is performed.

The corresponding entropy accretion rate is 
\begin{equation}
\left[\dot{\Xi}\right]_{\rm CF} =
\Lambda \frac{u\sqrt{1-\frac{2}{r}}}{\sqrt{1-u^2}} r^2 c_s^{\frac{2}{\gamma -1}}(\frac{\gamma -1}{\gamma - (1+c_s^2)})^\frac{1}{\gamma -1}
\label{anal58}
\end{equation}

The relationship between the corresponding space gradient of the speed of propagation of the acoustic perturbation
(the space gradient of the adiabatic sound speed) and that of the stationary advective
velocity can be found as

\begin{equation}
\left[\frac{dc_s}{dr}\right]_{\rm CF}=-\frac{\gamma -1}{2}\frac{(\frac{1}{u}+\frac{u}{1-u^2})\left[\frac{\text{du}}{\text{dr}}\right]_{\rm CF}+
\left\{\frac{2}{r}+\frac{1}{r^2(1-\frac{2}{r})}\right\}}{\frac{1}{c_s}+\frac{c_s}{\gamma -(1+c_s{}^2)}}
\label{anal59}
\end{equation}

The explicit expression for the velocity space gradient comes out to be
\begin{equation}
\left[\frac{du}{dr}\right]_{\rm CF} = \frac{c_s{}^2(\frac{2}{r}+\frac{1}{r^2(1-\frac{2}{r})})-f_2(r,\lambda )}{\frac{u}{1-u^2}(1-c_s{}^2)-
\frac{c_s{}^2}{u}}
\label{anal60}
\end{equation}

The critical point conditions are calculated as

\begin{equation}
\left|\left[u=c_s\right]_{r_c}, ~~~ \left[c_s\right]_{r_c}=\sqrt{\frac{f_2(r_c,\lambda )}{\frac{2}{r_c}+\frac{1}{r_c{}^2(1-\frac{2}{r_c})}}}\right|_{\rm CF}
\label{anal61}
\end{equation}

To compute the numerical value(s) of the critical point(s), one needs to substitute the critical point
condition into the expression for ${\cal E}$ to obtain an algebraic polynomial of the form
${\cal E} - f\left(r_c,\lambda,\gamma\right)=0$, which is
to be solved for $r_c$ for initial boundary conditions described by the astrophysically
relevant values of $\left[{\cal E},\lambda,\gamma\right]$. For conical
flow,
${\cal E} - f\left(r_c,\lambda,\gamma\right)=0$ provides a polynomial in $r_c$ of eleventh 
degree. The explicit expression for such polynomial is provided in the appendix. 

Polynomials of degree higher than four can not be solved
analytically. However, the number of roots of such equations lying between infinity and the
event horizons can be estimated analytically using the generalized Sturm sequence algorithm  \cite{shilpi}.

The
expressions for the critical values of $\left(du/dr\right)$ and
$\left(dc_s/dr\right)$ have been provided in the appendix.

The expression for the acoustic surface gravity can be obtained as
\begin{equation}
\kappa_{\rm Adiabatic}^{\rm Conical~Flow}=\left| \frac{r-2}{r^2 \left(1-c_s{}^2\right)}\sqrt{r^2-\lambda ^2 \left(1-\frac{2}{r}\right)} \left[\eta_1\frac{\text{du}}{\text{dr}}+\sigma_2\right] \right|_{\rm r_c}
\label{anal64a}
\end{equation}

where 
\begin{equation}
\sigma_2={\left(\frac{\gamma -1}{2}\right)\left(\frac{\frac{2}{r}+\frac{1}{r^2\left(1-\frac{2}{r}\right)}}{\frac{1}{u}+\frac{u}{\gamma -\left(1+u^2\right)}}\right)}
\label{sigma-CF}
\end{equation}

\subsubsection{Flow in hydrostatic equilibrium along the vertical direction}
\label{section6.1.3}
\noindent
The
radius dependent disc height is calculated as
\begin{equation}
H(r) = \frac{r^2c_s}{\lambda }\sqrt{\frac{2(1-u^2)(1-\frac{\lambda ^2}{r^2}(1-\frac{2}{r}))(\gamma -1)}{\gamma (1-\frac{2}{r})(\gamma -(1+c_s{}^2))}}
\label{anal64}
\end{equation}
and the corresponding mass accretion rate comes out to be
\begin{equation}
\dot{M}_{\rm VE}=4\pi \rho \frac{uc_sr^\frac{3}{2}}{\lambda }\sqrt{\frac{2(\gamma -1)(r^3-\lambda^2(r-2))}{\gamma(\gamma -(1+c_s{}^2))}}
\label{anal65}
\end{equation}

The entropy accretion rate is
\begin{equation}
\left[\dot{\Xi}\right]_{\rm VE}=
\sqrt{\frac{2}{\gamma }}\left[\frac{\gamma -1}{\gamma -(1+c_s{}^2)}\right]{}^{\frac{\gamma +1}{2(\gamma -1)}}\frac{c_s{}^{\frac{\gamma +1}{\gamma -1}}}{\lambda }\sqrt{1-\frac{\lambda ^2}{r^2}(1-\frac{2}{r})} \left( 4 \pi u r^3 \right)
\label{anal66}
\end{equation}

The relationship between $du/dr$ and $dc_s/dr$ is 
\begin{equation}
\left[\frac{\text{dc}_s}{\text{dr}}\right]_{\rm VE} =
\frac{-c_s\left\{\gamma -(1+c_s{}^2)\right\}}{\gamma +1}\left(\frac{1}{u}\left[\frac{\text{du}}{\text{dr}}\right]_{\rm VE}+
\frac{3}{r}+\frac{\lambda ^2}{r^3}\left\{\frac{1-\frac{3}{r}}{1-\frac{\lambda ^2}{r^2}(1-\frac{2}{r})}\right\}
\right)
\label{anal67}
\end{equation}

The explicit expression for $du/dr$ thus comes out to be
\begin{equation}
\left[\frac{du}{dr}\right]_{\rm VE} = \frac{\frac{2c_s{}^2}{\gamma +1}f_1(r,\lambda )-f_2(r,\lambda )}{\frac{u}{1-u^2}-\frac{2c_s{}^2}{(\gamma +1)u}}
\label{anal68}
\end{equation}

The critical point condition
\begin{equation}
\left| \left[u=\sqrt{\frac{1}{1+(\frac{\gamma +1}{2})(\frac{1}{c_s{}^2})}}\right]_{r_c}
=\sqrt{\frac{f_2(r_c,\lambda )}{f_1(r_c,\lambda )+f_2(r_c,\lambda)}} \right|_{\rm VE}
\label{anal69}
\end{equation}
indicates that critical surfaces are not the acoustic horizons for flow in vertical equilibrium since the value of the
Mach number $M$ at the
critical point is found to be
\begin{equation}
M_c=
\sqrt{
\left({\frac{2}{\gamma+1}}\right)
\frac
{{f_{1}}(r_c,\lambda)}
{{{f_{1}}(r_c,\lambda)}+{{f_{2}}(r_c,\lambda)}}
}
\label{anal69a}
\end{equation}

Unlike other flow models, to locate the radius of the acoustic horizon $r_h$ for
flow in vertical equilibrium, one needs to integrate the flow equations from the critical point upto the
radial distance where Mach number becomes unity.

The location of the critical point can be obtained by solving an $8^{\rm th}$ degree polynomial equation,
explicit expression of which is provided in the appendix. The expression for critical velocity gradients 
are also provided in the appendix. 

The acoustic surface gravity can be calculated as 
\begin{equation}
\kappa_{\rm Adiabatic}^{\rm Vertical~Equilibrium}=\left| \frac{r-2}{r^2 \left(1-c_s{}^2\right)}\sqrt{r^2-\lambda ^2 \left(1-\frac{2}{r}\right)} \left[\eta_3\frac{\text{du}}{\text{dr}}+\sigma_3\right] \right|_{\rm r_s}
\label{anal72a}
\end{equation}
where $r_s$ is the sonic point which is to be obtained by integrating the flow equations from
the corresponding critical points, and
\begin{equation}
\eta_3 = {1+\frac{c_s}{u}\left(\frac{\gamma -\left(1+c_s{}^2\right)}{\gamma +1}\right)},~~
\sigma_3={\frac{c_sf_1\left(\gamma -\left(1+c_s{}^2\right)\right)}{\gamma +1}}
\label{eta-sigma-VE}
\end{equation}

\subsection{Isothermal accretion}
\label{section6.2}
As is understood, the expression for the mass accretion rate for constant height and 
conical flow model for isothermal accretion will exactly be the same as those 
obtained for the adiabatic flow since the flow thickness does not depend on $\gamma$. For 
accretion in vertical equilibrium, however, the flow thickness will be different for two 
different equation of states and hence the corresponding mass accretion rate for the isothermal 
flow will also be different from its polytropic counterpart
\begin{equation}
\left[\dot M\right]^{\rm Isothermal}_{\rm VE}=4\pi \rho \frac{r^\frac{3}{2} u c_s}{\lambda} \sqrt{2(r^3-(r-2 ) \lambda ^2)}
\label{anal88}
\end{equation}
The corresponding (advective) velocity (space) gradients are obtained as
\begin{equation}
\left[\frac{du}{dr}\right]_{\rm CH}^{\rm Iosthermal}=\frac{\left(2r^3-2 (r-2)^2 \lambda^2+(1-r) \left(2r^3+4 \lambda^2-2 r \lambda^2\right) c_s^2\right)u(u^2-1)}{(2 -r) r \left(-2r^3-4 \lambda^2+2 r \lambda^2\right) \left(u^2-c_s^2\right)}
\label{anal76}
\end{equation}
\begin{equation}
\left[\frac{du}{dr}\right]_{\rm CF}^{\rm Isothermal}=
\frac{\left\{ 2r^3-2(r-2)^2\lambda ^2+(3-2r)\left(2r^3+4 \lambda ^2-2r \lambda ^2\right)c_s^2\right\}u(u^2-1)}{(2-r)r\left(-2r^3-4 \lambda ^2+2r \lambda ^2\right)(u^2-c_s^2)}
\label{anal82}
\end{equation}
\begin{equation}
\left[\frac{du}{dr}\right]_{\rm VE}^{\rm Isothermal}=\frac{\left[r^3-(r-2)^2 \lambda ^2 +(2 -r) (3 r^3+3 \lambda ^2-2 r \lambda ^2 ) c_s^2 \right] u (u^2-1)}{\frac{1}{2} r (r-2 ) \left(-2r^3-4 \lambda ^2+2 r \lambda ^2 \right) \left[c_s^2-\left(1+c_s^2\right) u^2\right]}
\label{anal89}
\end{equation}
In the appendix, we describe how one can obtain the critical point condition, the 
location of the sonic horizon(s), and the value of the respective quantities (required to 
estimate $\kappa$ for isothermal flow) evaluated on such horizons. 

A two parameter set $\left[T,\lambda\right]$ is to be provided to calculate the location of the 
critical points {\it completely analytically} (unlike the polytropic flow where the number of 
critical points could be computed analytically but not their locations). One interesting point to 
note is that unlike the accretion flow under the post Newtonian pseudo-Schwarzschild 
potentials (as derived in \cite{nag12}), for complete general relativistic flow the critical surface 
does not coincide with the acoustic horizon for flow in vertical equilibrium even for isothermal 
accretion. 
Using the values of $\left[u,\frac{du}{dr}\right]_{\rm r_h}$\footnote{For isothermal 
flow, the sound speed remains constant for all values of 
$2<r<\infty$.} as derived in the appendix, we derive the
corresponding expressions for the acoustic surface gravities for three different flow models for isothermal 
accretion
\begin{equation}
\kappa_{\rm Isothermal}^{\rm Constant~Height}=\left| \frac{r-2}{r^2 \left(1-c_s{}^2\right)}\sqrt{r^2-\lambda ^2 \left(1-\frac{2}{r}\right)}\left(\frac{\text{du}}{\text{dr}}\right) \right|_{\rm r_c}
\label{anal80a}
\end{equation}
\begin{equation}
\kappa_{\rm Isothermal}^{\rm Conical~Flow}=\left| \frac{r-2}{r^2 \left(1-c_s{}^2\right)}\sqrt{r^2-\lambda ^2 \left(1-\frac{2}{r}\right)}\left(\frac{\text{du}}{\text{dr}}\right) \right|_{\rm r_c}
\label{anal86a}
\end{equation}
\begin{equation}
\kappa_{\rm Isothermal}^{\rm Vertical~Equilibrium}=\left| \frac{r-2}{r^2 \left(1-c_s{}^2\right)}\sqrt{r^2-\lambda ^2 \left(1-\frac{2}{r}\right)}\left(\frac{\text{du}}{\text{dr}}\right) \right|_{\rm r_s}
\label{anal94}
\end{equation}
where $r_s$ is the sonic point which is to be obtained by integrating the flow equations from 
the corresponding critical point. 

\section{Dependence of acoustic surface gravity on flow geometry -- Polytropic accretion}
\label{dependence-on-polytropic}
\noindent
In this section we will manifest how the geometric configuration of the
non self-gravitating axially symmetric stationary background flow in a Schwarzschild 
metric influences the computation of the acoustic surface gravity for 
polytropic accretion onto astrophysical black holes. 
We first construct the parameter space determined by the initial boundary conditions 
to show the parameter dependence of the multi-critical flow behavior, and then 
will describe how one can pick certain regions of $\left[{\cal E},\lambda,\gamma\right]$
space for which mono-transonic accretion is possible for all three different geometric configurations. 
\subsection{The parameter space classification}
\label{parameter-space-polytropic}
\noindent
The critical point(s) are 
obtained once $\left[{\cal E},\lambda,\gamma\right]$ is specified. A three 
dimensional parameter space spanned by $\left[{\cal E},\lambda,\gamma\right]$ and bounded 
by an astrophysically relevant range 
$\left[1<{\cal E}<2,0<\lambda{\le}4,4/3{\le}\gamma{\le}5/3\right]$, can thus be 
explored to understand the dependence of the multi-critical behavior on initial 
boundary conditions. 

For the sake of convenience, a two dimensional projection of such a three dimensional 
parameter space will be analyzed. $^3{C}_2$ allowed combinations of such projections
are available. In this work, we prefer to project the 
$\left[{\cal E},\lambda,\gamma\right]$
space on a $\left[{\cal E},\lambda\right]$ plane by keeping the 
adiabatic constant fixed to the value $\gamma=4/3$. Although such  $\left[{\cal E},\lambda\right]$
projections can be studied for any other values lying in the range $4/3{\le}\gamma{\le}5/3$
as well. 

In figure 1, we study the ${\cal E} - \lambda$ plane for three different flow geometries. 
Variation of ${\cal E} - \lambda$ branches for flow in hydrostatic equilibrium
along the vertical direction, conical flow and flow with constant thickness
are represented by solid red lines, dashed green lines, and dotted blue lines, respectively.
Hereafter, we will follow the aforementioned color scheme to show results corresponding to the
flow geometries discussed above.
For flow with constant height, A$_1$A$_2$A$_3$A$_4$ represents the region of 
$\left[{\cal E},\lambda\right]$ for which eq. (\ref{anal42}) along with the corresponding critical point conditions provides three real 
positive roots lying outside the gravitational horizon. For region A$_1$A$_2$A$_3$, 
one finds ${\dot {\Xi}}_{\rm inner} > {\dot {\Xi}}_{\rm outer}$ and accretion is 
multi-critical. A subspace of A$_1$A$_2$A$_3$ allows shock formation. Such a
subspace provides true multi-transonic accretion where the stationary transonic 
solution passing through the outer sonic point joins with the stationary transonic 
solution constructed through the inner sonic point through a discontinuous energy preserving shock
of Rankine-Hugoniot type. Such shocked multi-transonic solution 
contains two smooth transonic (from sub to super) transitions at two regular 
sonic points (of saddle type) and a discontinuous transition (from super to sub)
at the shock location. 

On the other hand, the region A$_1$A$_3$A$_4$ represents the subset of 
$\left[{\cal E},\lambda,\gamma\right]_{\rm mc}$ (where 
`mc' stands for `multi-critical') for which 
${\dot {\Xi}}_{\rm inner} < {\dot {\Xi}}_{\rm outer}$ and hence 
incoming flow can have only one critical point of saddle type and the background flow
possesses one acoustic horizon at the inner saddle type sonic point. 
The boundary A$_1$A$_3$ between these two regions represents the value of $\left[{\cal E},\lambda,\gamma\right]$
for which multi-critical accretion is characterized by 
${\dot {\Xi}}_{\rm inner} = {\dot {\Xi}}_{\rm outer}$ and hence the 
transonic solutions passing through the inner and the outer sonic 
points are completely degenerate, leading to the formation of a 
heteroclinic orbit \footnote{Heteroclinic orbits are the trajectories defined on a phase
portrait which connects two different saddle type critical points. Integral
solution configuration on phase portrait characterized by heteroclinic
orbits are topologically unstable \cite{js99,diff-eqn-book,strogatz}.}
on the phase portrait. Such flow configuration 
can not be used to study the analogue properties as we believe since 
it does not have uniqueness in forming the acoustic horizons. Such flow pattern 
is subjected to instability and turbulence as well \cite{das-czerny-2012-new-astronomy}.

\begin{SCfigure}
\centering
\includegraphics[scale=0.675,angle=0.0]{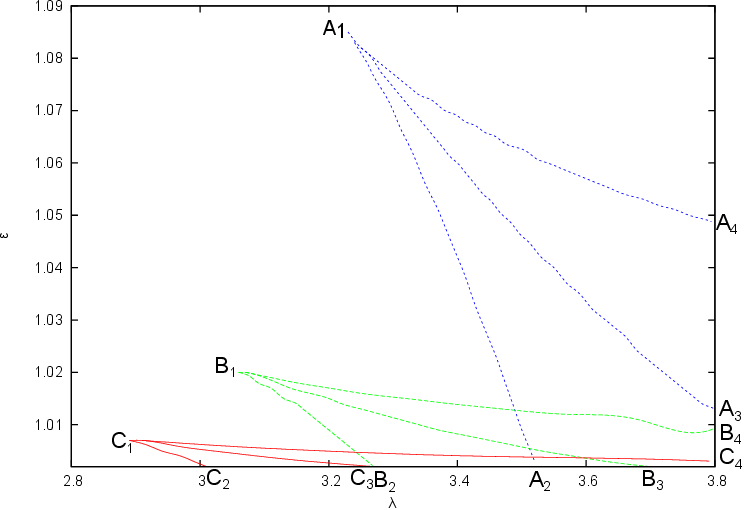}
\caption[]{${\cal E} - \lambda$ plane for three different flow geometries for adiabatic accretion
for fixed value of $\gamma=4/3$. 
Variation of ${\cal E} - \lambda$ branches for flow in hydrostatic equilibrium
along the vertical direction, conical flow and flow with constant thickness
are represented by solid red lines, dashed green lines, and dotted blue lines, respectively.
See section \ref{parameter-space-polytropic} for further details about the 
parameter space classification.}
\label{fig1}
\end{SCfigure}

\begin{SCfigure}
\centering
\includegraphics[scale=0.635,angle=0.0]{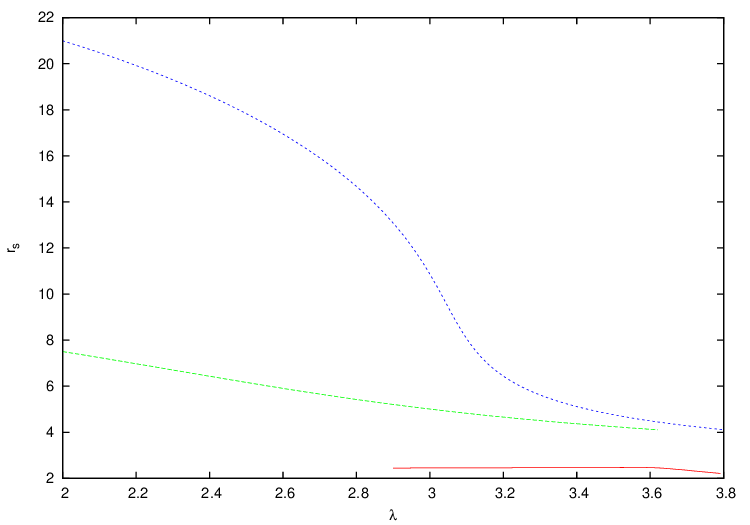}
\caption[]{For $\left[{\cal E}=1.12,\gamma=4/3\right]$, variation of the location of the acoustic horizons as a function of the flow 
angular momentum $\lambda$ for stationary mono-transonic adiabatic accretion passing through the 
inner sonic point $r_s$ for the three different geometric configurations of the flow 
considered in this work. $r_s - \lambda$ curves for flow in hydrostatic equilibrium
along the vertical direction, conical flow and flow with constant thickness
are represented by solid red lines, dashed green lines, and dotted blue lines, respectively.
}
\label{fig2}
\end{SCfigure}

\begin{SCfigure}
\centering
\includegraphics[scale=0.635,angle=0.0]{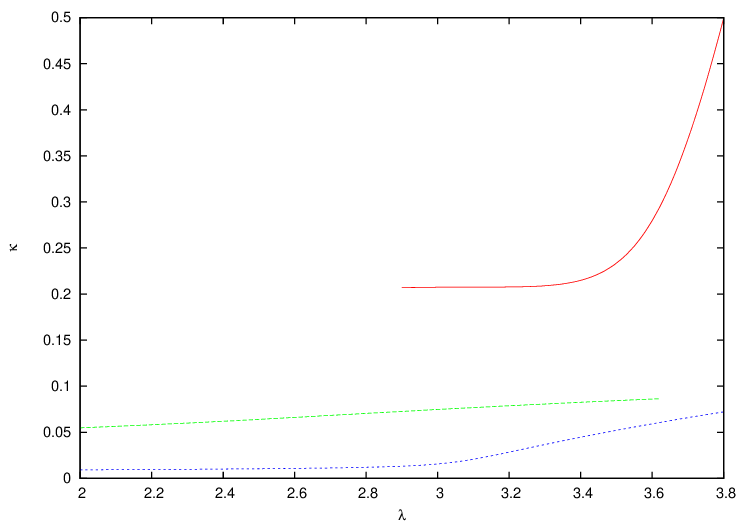}
\caption[]{For the same initial boundary conditions used to obtain figure \ref{fig2},
variation of acoustic surface gravity $\kappa$ with the flow
angular momentum $\lambda$ for stationary mono-transonic adiabatic accretion passing through the
inner sonic point $r_s$ for the three different geometric configurations of the flow
considered in this work. $\kappa - \lambda$ curves for flow in hydrostatic equilibrium
along the vertical direction, conical flow and flow with constant thickness
are represented by solid red lines, dashed green lines, and dotted blue lines, respectively.
}
\label{fig3}
\end{SCfigure}

\begin{SCfigure}
\centering
\includegraphics[scale=0.635,angle=0.0]{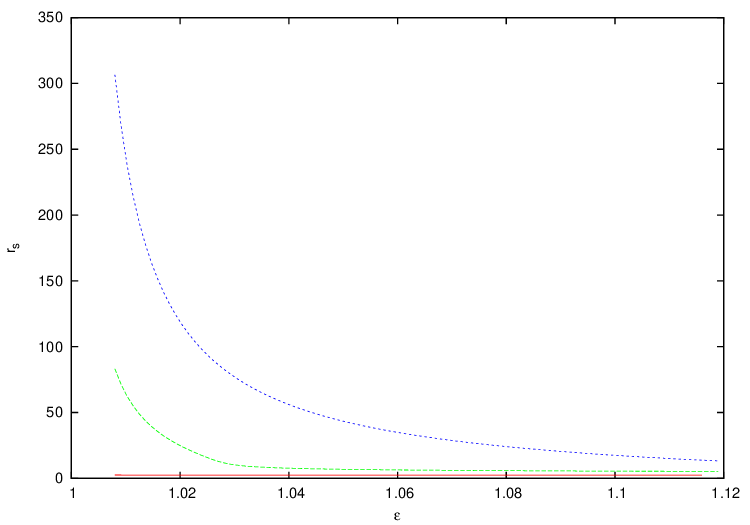}
\caption[]{For $\left[\lambda=2.89,\gamma=4/3\right]$, 
variation of the location of the acoustic horizons as a function of the 
specific energy of the flow ${\cal E}$
for stationary mono-transonic adiabatic accretion passing through the
inner sonic point $r_s$ for the three different geometric configurations of the flow
considered in this work. $r_s - \lambda$ curves for flow in hydrostatic equilibrium
along the vertical direction, conical flow and flow with constant thickness
are represented by solid red lines, dashed green lines, and dotted blue lines, respectively.
}
\label{fig4}
\end{SCfigure}


\begin{SCfigure}
\centering
\includegraphics[scale=0.635,angle=0.0]{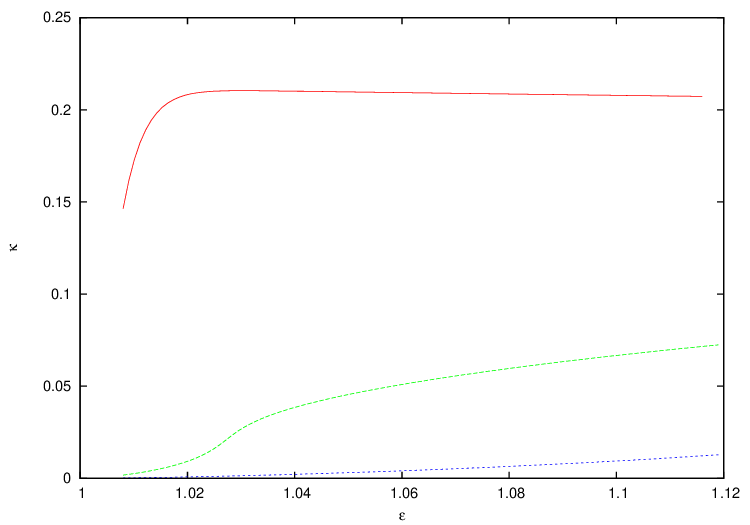}
\caption[]{For the same initial boundary conditions used to obtain figure \ref{fig4},
variation of acoustic surface gravity $\kappa$ with the specific energy of the flow ${\cal E}$
for stationary mono-transonic adiabatic accretion passing through the
inner sonic point $r_s$ for the three different geometric configurations of the flow
considered in this work. $\kappa - {\cal E}$ curves for flow in hydrostatic equilibrium
along the vertical direction, conical flow and flow with constant thickness
are represented by solid red lines, dashed green lines, and dotted blue lines, respectively.
}
\label{fig5}
\end{SCfigure}

\begin{SCfigure}
\centering
\includegraphics[scale=0.635,angle=0.0]{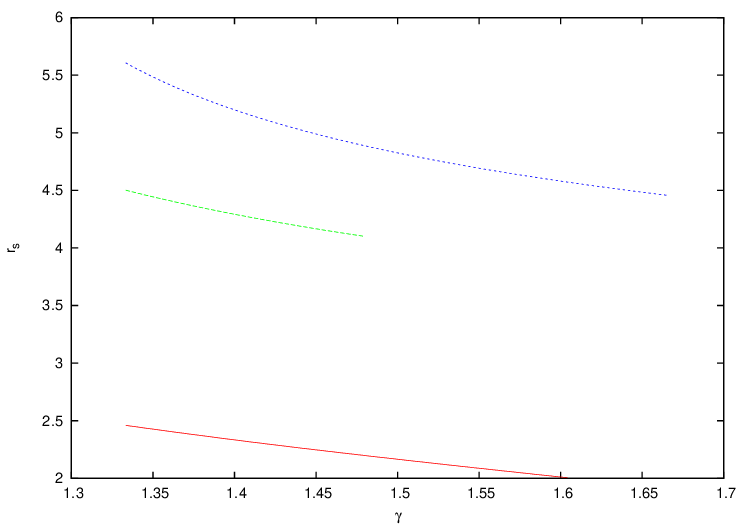}
\caption[]{For $\left[{\cal E}=1.12,\lambda=3.3\right]$, 
variation of the location of the acoustic horizons as a function of the 
adiabatic index $\gamma$
for stationary mono-transonic adiabatic accretion passing through the
inner sonic point $r_s$ for the three different geometric configurations of the flow
considered in this work. $r_s - \gamma$ curves for flow in hydrostatic equilibrium
along the vertical direction, conical flow and flow with constant thickness
are represented by solid red lines, dashed green lines, and dotted blue lines, respectively.
}
\label{fig6}
\end{SCfigure}


\begin{SCfigure}
\centering
\includegraphics[scale=0.635,angle=0.0]{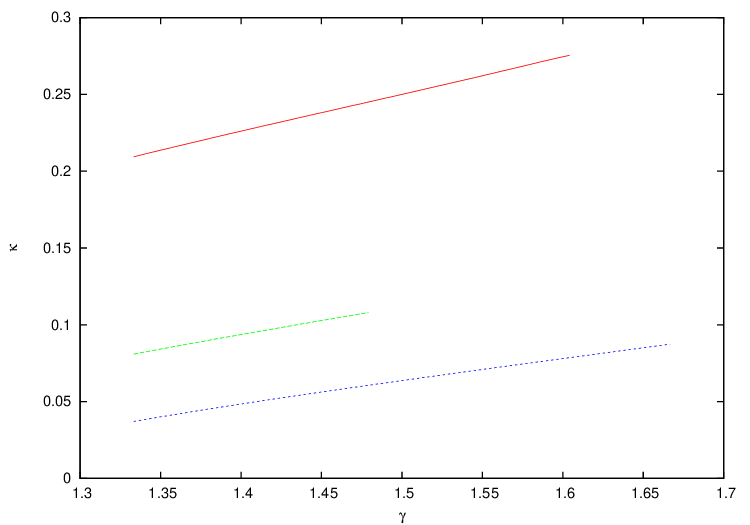}
\caption[]{For the same initial boundary conditions used to obtain figure \ref{fig6},
variation of acoustic surface gravity $\kappa$ with the adiabatic index $\gamma$ has been plotted
for stationary mono-transonic adiabatic accretion passing through the
inner sonic point $r_s$ for the three different geometric configurations of the flow
considered in this work. $\kappa - \gamma$ curves for flow in hydrostatic equilibrium
along the vertical direction, conical flow and flow with constant thickness
are represented by solid red lines, dashed green lines, and dotted blue lines, respectively.
}
\label{fig7}
\end{SCfigure}


Similar analysis can be carried out for other two flow geometries.
B$_1$B$_2$B$_3$B$_4$ and C$_1$C$_2$C$_3$C$_4$ represents 
$\left[{\cal E},\lambda,\gamma\right]_{\rm mc}{\subset}\left[{\cal E},\lambda,\gamma\right]$
for the quasi-spherical conical flow and for flow in vertical equilibrium, respectively. 
For B$_1$B$_2$B$_3$B$_4$, B$_1$B$_2$B$_3$ represents the multi-critical flow 
with ${\dot {\Xi}}_{\rm inner} > {\dot {\Xi}}_{\rm outer}$ and B$_1$B$_3$B$_4$
represents such region for ${\dot {\Xi}}_{\rm inner} < {\dot {\Xi}}_{\rm outer}$, 
B$_1$B$_3$ being the interface between them representing the values of 
$\left[{\cal E},\lambda,\gamma\right]$ for which only the heteroclinic connections 
are obtained. Similar classifications can also be made for the region C$_1$C$_2$C$_3$C$_4$
as well. 

We would like to pick a range of $\left[{\cal E},\lambda,\gamma\right]$
for which all three flow configurations will provide mono-transonic accretion. Moreover, we are 
interested mainly in the stationary mono-transonic solutions passing though the inner 
sonic point since the acoustic surface gravity evaluated at the inner acoustic 
horizon is of the order of magnitude (upto about 10$^5$ or even higher)
higher than the acoustic surface gravity evaluated at the outer acoustic 
horizon. In addition, the location of the outer acoustic horizon, and hence 
the value of the acoustic surface gravity evaluated on it, are not much sensitive 
to the variation of $\left[{\cal E},\lambda,\gamma\right]$
compared to their counterparts corresponding to the inner acoustic horizon. 


Out of the three parameters ${\cal E},\lambda$ and $\gamma$, we choose one parameter, say $\lambda$,
to vary by keeping the values of $\left[{\cal E},\gamma\right]$ fixed, for three different 
flow configurations for the mono-transonic flow through the inner acoustic horizon
to obtain the $\left[\kappa - \lambda\right]$ variation. Three different $\left[\kappa - \lambda\right]$
variations for three different flow geometries will then be compared to examine the influence of the 
flow configuration on the value of the acoustic surface gravity. 
We perform the same operation for the other two parameters ${\cal E}$ and $\lambda$, to obtain 
the $\left[\kappa - {\cal E}\right]$ and $\left[\kappa - \gamma\right]$ variations 
for three different flow geometries by keeping $\left[\lambda,\gamma\right]$ and 
$\left[{\cal E},\lambda\right]$ invariant, respectively. Finally, we perform 
similar exercise for isothermal accretion of three different flow geometries to obtain and to 
mutually compare the $\left[\kappa - \lambda\right]$ and $\left[\kappa - T\right]$ variations,
respectively. 

\subsection{Variation of $\kappa$ with $\left[{\cal E},\lambda,\gamma\right]$}
\label{kappa-variation-polytropic}
\noindent
In figure 2, for a fixed set of $\left[{\cal E}=1.12,\gamma=4/3\right]$\footnote{Such a high value
of ${\cal E}$ (`hot' accretion) has been considered to ensure that
mono-transonic stationary solution passing through the inner sonic point
is obtained for all three different geometrical configurations of the axisymmetric matter
considered in this work.}, we plot the
location of the inner type acoustic horizon (the inner sonic point $r_s$) as a
function of the specific angular momentum $\lambda$ of the flow for mono-transonic
stationary
accretion solution for three different flow geometries. 
It is observed that the location of the acoustic horizon anti-correlates with 
$\lambda$. This is somewhat
obvious because for greater amount of rotational energy content of the flow,
accretion starts with smaller advective velocity and has to approach very close
to the event horizon to acquire the dynamical velocity sufficiently
large to smoothly overcome the acoustic velocity.

For a specified initial boundary condition 
describing the flow, one observes
\begin{equation}
r_{s}^{\rm vertical} < r_{s}^{\rm conical} < r_{s}^{\rm constant~height}
\label{anal95}
\end{equation}
This indicates that for the same set of $\left[{\cal E},\lambda,\gamma\right]$, the
acoustic horizon for accretion in hydrostatic equilibrium along the vertical direction
forms at the closest proximity of the black hole event horizon and hence the
relativistic acoustic geometry at the neighborhood of such acoustic horizons are
subjected to considerably strong gravity space time. One thus intuitively concludes that
among all three flow configurations considered in this work,
the Hawking like effects may perhaps be more pronounced for axisymmetric background flow
in hydrostatic equilibrium along the vertical direction. This intuitive conclusion is
further supported by results represented in figure 3 where we have studied the variation of the
acoustic surface gravity $\kappa$ as a function of the flow angular momentum $\lambda$
for the same set of initial boundary conditions as well as for the span of $\lambda$ for which
figure 2 has been obtained. For identical initial boundary conditions as determined by
$\left[{\cal E},\lambda,\gamma\right]$, one obtains
\begin{equation}
\kappa^{\rm vertical} > \kappa^{\rm conical} > \kappa^{\rm constant~height}
\label{anal96}
\end{equation}
For a fixed value of $\left[\lambda=2.89,\gamma=4/3\right]$, the variation of the
location of the inner acoustic horizons (the inner sonic points $r_s$) as a
function of the specific flow energy ${\cal E}$ is plotted in figure 4. $r_s$ anti-co-relates
with ${\cal E}$ for obvious reasons. At a large distance
away from the accretor, the total specific energy is essentially determined by the thermal
energy of the flow, a large value of ${\cal E}$ (`hot' accretion) corresponds to a
high value of the sound speed $c_s$ to begin with. The subsonic to the
supersonic transition takes place quite close to the black hole where the bulk flow
velocity (the advective velocity $u$) becomes large enough to overcome the sound speed.
Once again, flow in vertical equilibrium produces the acoustic horizons located at
a relatively stronger gravity region. The analogue effect should be more
pronounced for such geometric configuration of the flow. Results presented in figure 5, where
the acoustic surface gravity $\kappa$ has been plotted as a function of the specific energy of
the flow for the same set of $\left[\lambda,\gamma\right]$ used to draw figure 4, asserts such conclusion.
For a fixed value of $\left[{\cal E}= 1.12,\lambda=3.3\right]$, in figure 6
we find that the location of the acoustic horizon anti-correlates with $\gamma$ and hence 
the acoustic surface gravity $\kappa$ co-relates with $\gamma$ as expected (as shown in figure 7). 
Similar result can be obtained for any set of $\left[{\cal E},\lambda\right]$ for which 
monotransonic stationary accretion passing through the inner type sonic point 
can be obtained for all three flow configurations considered in this work.
Here too the acoustic surface gravity
for accretion in hydrostatic equilibrium along the vertical direction is maximum (compared to 
the conical flow and flow with constant thickness) for the same set of initial boundary conditions.
We thus obtain 
\begin{eqnarray}
r_{s}^{\rm vertical} < r_{s}^{\rm conical} < r_{s}^{\rm constant~height}\\
\nonumber
\kappa^{\rm vertical} > \kappa^{\rm conical} > \kappa^{\rm constant~height}
\label{anal96a}
\end{eqnarray}
here as well.


\section{Dependence of acoustic surface gravity on flow geometry -- Isothermal accretion}
\label{dependence-on-isothermal}
\noindent 

The $\left[T,\lambda\right]$ 
parameter space will first be constructed to manifest the 
multi-critical flow behavior. A certain subset of the entire $\left[T,\lambda\right]$
will then be chosen for which isothermal accretion in all three matter geometries will 
have mono-transonic solutions constructed through the inner sonic point. 
\subsection{The parameter space classification}
\label{parameter-space-isothermal}
\noindent
Figure 8 depicts the 
parameter space division labeled following the scheme introduced in 
section \ref{parameter-space-polytropic}. For constant height flow,
conical flow and vertical equilibrium flow, 
A${^\prime}{_1}$A${^\prime}{_2}$A${^\prime}{_3}$A${^\prime}{_4}$,
B${^\prime}{_1}$B${^\prime}{_2}$B${^\prime}{_3}$B${^\prime}{_4}$,
and C${^\prime}{_1}$C${^\prime}{_2}$C${^\prime}{_3}$C${^\prime}{_4}$,
represents the $\left[T,\lambda\right]$ regions for which 
eq. (\ref{apen20}), eq. (\ref{apen22}) and eq. (\ref{apen24}) will 
provide three real physical roots located outside the gravitational horizon,
respectively. It is to be noted that 
figure 8 can be obtained 
completely analytically solving the 
representative equations using the Ferrari's method 
\cite{kurosh72}. 

Similar to the polytropic accretion, the wedge shaped region 
$\left[T,\lambda\right]_{\rm mc}$, where `mc' stands for 
`multi-critical', has two subsections divided by a distinct 
boundary. However, unlike the entropy accretion rate 
${\dot {\Xi}}$ for the polytropic flow, the first integral of 
motion $\xi$ as defined in eq. (\ref{anal73}) determines the 
characteristic features of various subspaces of 
$\left[T,\lambda\right]_{\rm mc}$. For constant height flow,
A${^\prime}{_1}$A${^\prime}{_2}$A${^\prime}{_3}$A${^\prime}{_4}$
region is subdivided into A${^\prime}{_1}$A${^\prime}{_2}$A${^\prime}{_3}$
for which $\xi_{\rm in} > \xi_{\rm out}$ and 
A${^\prime}{_1}$A${^\prime}{_3}$A${^\prime}{_4}$ for which $\xi_{\rm in} < \xi_{\rm out}$ with the 
boundary line A${^\prime}{_1}$A${^\prime}{_3}$ on which $\xi_{\rm in} = \xi_{\rm out}$.
$\left[T,\lambda\right]_{\rm A{^\prime}{_1}A{^\prime}{_2}A{^\prime}{_3}}$ provides the 
multi-critical integral solutions for which the 
homoclinic orbit is constructed through the inner critical point whereas
$\left[T,\lambda\right]_{\rm A{^\prime}{_1}A{^\prime}{_3}A{^\prime}{_4}}$ produces the 
multi-critical solution for which accretion is mono transonic and the 
corresponding homoclinic orbit is constructed through the outer 
critical point. $\left[T,\lambda\right]_{\rm A{^\prime}{_1}A{^\prime}{_3}}$ 
provides the heteroclinic orbits for which one obtains degenerate accretion 
solutions. 

\begin{SCfigure}
\centering
\includegraphics[scale=0.39,angle=0.0]{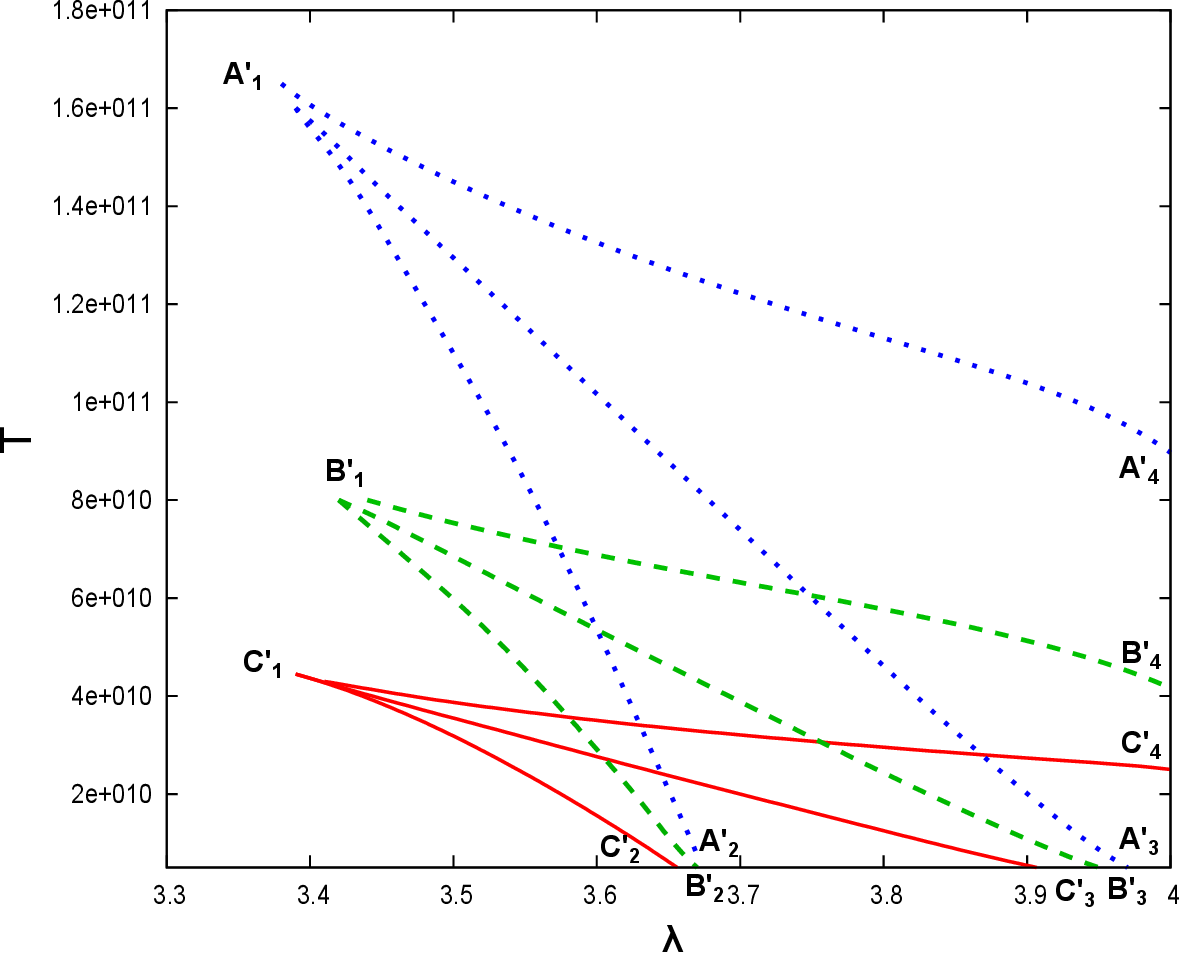}
\caption[]{$T - \lambda$ plane for three different flow geometries for isothermal accretion.
Variation of $T - \lambda$ branches for flow in hydrostatic equilibrium
along the vertical direction, conical flow and flow with constant thickness
are represented by solid red lines, dashed green lines, and dotted blue lines, respectively.
See section \ref{parameter-space-isothermal} for further details about the
parameter space classification.}
\label{fig8}
\end{SCfigure}

For a certain subset of $\left[T,\lambda\right]_{\rm A{^\prime}{_1}A{^\prime}{_2}A{^\prime}{_3}}$
$\in$ $\left[T,\lambda\right]_{\rm mc}$, temperature preserving shock may 
form to provide true multi-transonicity. Such stationary solutions contain 
two acoustic black hole horizons at the inner and the outer sonic points and 
an acoustic white hole solution at the shock location. We, however, will not 
perform the shock finding analysis in the present work. 

In a similar spirit, subdivisions in multi-critical parameter spaces 
for the conical and the vertical equilibrium can also be obtained.


\begin{SCfigure}
\centering
\includegraphics[scale=0.635,angle=0.0]{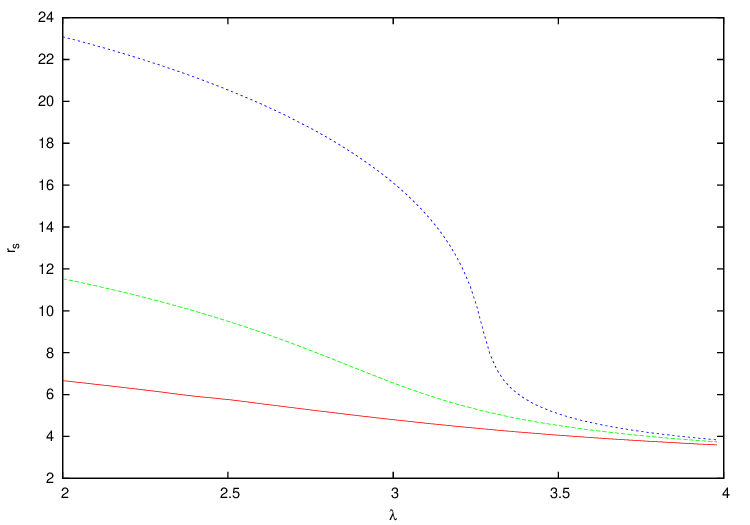}
\caption[]{For $T_{10}=20.0$ (in units of $10^10$ K), variation of the location of the acoustic horizons as a function of the flow
angular momentum $\lambda$ for stationary mono-transonic isothermal accretion passing through the
inner sonic point $r_s$ for the three different geometric configurations of flow
considered in this work. $r_s - \lambda$ curves for flow in hydrostatic equilibrium
along the vertical direction, conical flow and flow with constant thickness
are represented by solid red lines, dashed green lines, and dotted blue lines, respectively.
}
\label{fig9}
\end{SCfigure}

\begin{SCfigure}
\centering
\includegraphics[scale=0.635,angle=0.0]{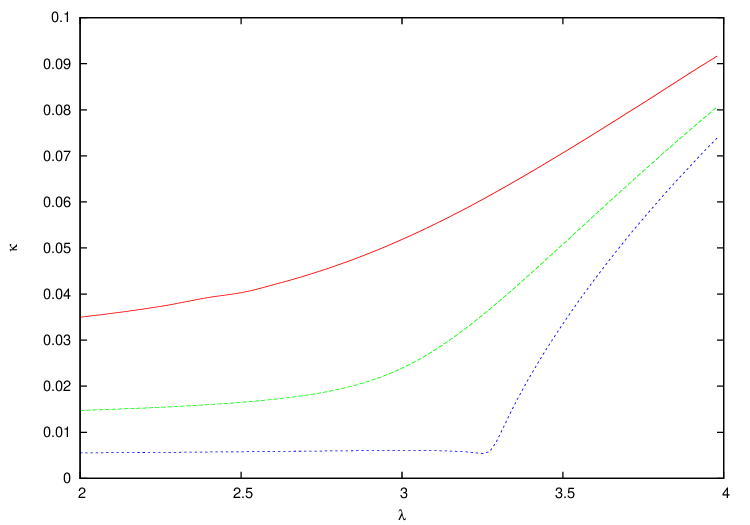}
\caption[]{For the same initial boundary conditions used to obtain figure \ref{fig9},
variation of acoustic surface gravity $\kappa$ with the flow
angular momentum $\lambda$ for stationary mono-transonic isothermal accretion passing through the
inner sonic point $r_s$ for the three different geometric configurations of flow
considered in this work. $\kappa - \lambda$ curves for flow in hydrostatic equilibrium
along the vertical direction, conical flow and flow with constant thickness
are represented by solid red lines, dashed green lines, and dotted blue lines, respectively.
}
\label{fig10}
\end{SCfigure}


\begin{SCfigure}
\centering
\includegraphics[scale=0.635,angle=0.0]{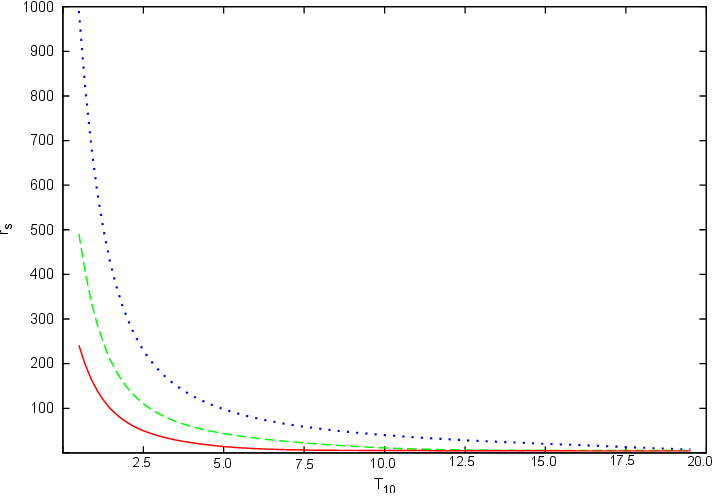}
\caption[]{For $\lambda=3.3$,
variation of the location of the acoustic horizon with the constant flow temperature T
(in units of $10^{10}$ Kelvin and denoted as $T_{10}$)
for stationary mono-transonic isothermal accretion passing through the
inner sonic point $r_s$ for the three different geometric configurations of the flow
considered in this work. $r_s - T_{10}$ curves for flow in hydrostatic equilibrium
along the vertical direction, conical flow and flow with constant thickness
are represented by solid red lines, dashed green lines, and dotted blue lines, respectively.
}
\label{fig11}
\end{SCfigure}

\begin{SCfigure}
\centering
\includegraphics[scale=0.635,angle=0.0]{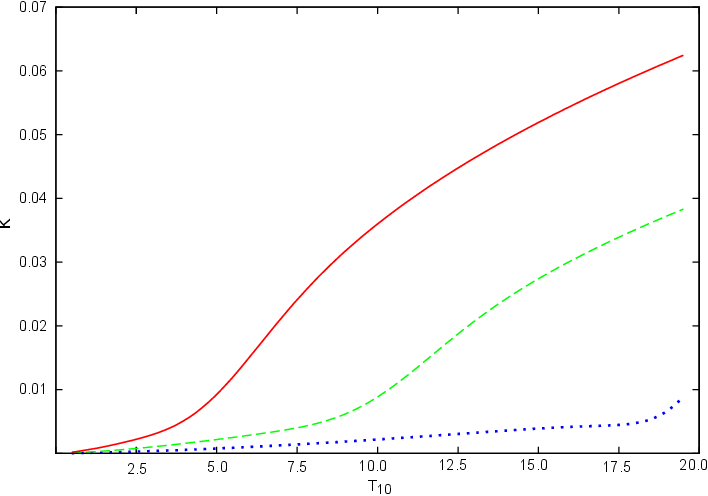}
\caption[]{For the same initial boundary conditions used to obtain figure \ref{fig11},
variation of acoustic surface gravity $\kappa$ with the 
constant flow temperature T
(in units of $10^{10}$ Kelvin and denoted as $T_{10}$)
for stationary mono-transonic isothermal accretion passing through the
inner sonic point $r_s$ for three different geometric configuration of the flow
considered in this work. $\kappa - T_{10}$ curves for flow in hydrostatic equilibrium
along the vertical direction, conical flow and flow with constant thickness
are represented by solid red lines, dashed green lines, and dotted blue lines, respectively.
}
\label{fig12}
\end{SCfigure}

\subsection{Variation of $\kappa$ with $\left[\lambda,T\right]$}
\noindent
In figure 9, we plot the variation of the location of the acoustic horizon (the inner sonic
point $r_s$ with the constant specific angular momentum of the flow $\lambda$. Tha value of flow temperature $T$ expressed in units of $10^10$ K and denoted by $T_{10}$ has been fixed at 20.0, thus ensuring mono-transonic accretion for all three flow geometries. For
obvious reasons (as described in section \ref{kappa-variation-polytropic}), $r_s$ anti-correlates with
$\lambda$. For accretion flow in hydrostatic equilibrium along the vertical direction, entire range
of sonic points produced (for the domain of $\lambda$ considered in this work) lie in the 
close proximity of the event horizon. This indicates that for the same set of
initial boundary conditions describing the flow, the Hawking like effects will be
maximally pronounced for such a flow model. Such a conclusion is further reinforced from
results presented in figure 10 where we have plotted the acoustic surface gravity $\kappa$ as
a function of the flow angular momentum $\lambda$ to obtain
\begin{equation}
\kappa^{\rm vertical} > \kappa^{\rm conical} > \kappa^{\rm constant~height}
\label{anal97}
\end{equation}

Figure 11 represents the variation of the location of the acoustic horizon with
constant flow temperature in units of $T_{10}$. Since the position independent sound speed ${c_s}{\propto}T^{\frac{1}{2}}$,
the sonic point $r_s$ anti-correlates with the flow temperature $T$. In figure 12 we plot the
variation of the acoustic surface gravity $\kappa$ as a function of $T$ in units of $T_{10}$.
Hotter flow produces larger value of the acoustic surface gravity $\kappa$. 


\section{Concluding remarks}
\label{concluding-remarks}
\noindent
For analogue models, the mass of the system itself can not determine
the value of acoustic surface gravity (as well as the 
associated Hawking like temperature) since $\kappa$ is obtained as 
a complicated non linear functional\footnote{$\kappa$ is actually 
functions of  $\left[u,c_s,du/dr,dc_s/dr\right]$ for adiabatic flow 
and $\left[u,c_s,du/dr\right]$ of isothermal flow. $\left[u,c_s,du/dr,dc_s/dr\right]_{\rm adia}$
and $\left[u,c_s,du/dr\right]_{\rm iso}$ are non linear functions of $\left[{\cal E},\lambda,\gamma\right]_{\rm adia}$
and $\left[T,\lambda\right]_{\rm iso}$ respectively, and hence 
$\kappa_{\rm adia}$ and $\kappa_{\rm iso}$ are functionals of 
$\left[{{\cal E},\lambda,\gamma}\right]_{\rm adia}$ and 
$\left[T,{\lambda}\right]_{\rm iso}$, respectively.} of
the initial boundary conditions describing the flow profile. The same set of 
initial boundary conditions may provide significantly different 
phase portrait for integral stationary solutions for different geometric 
configurations of the background matter flow, and it is necessary to study the 
influence of the matter geometry on the determination of the corresponding 
relativistic acoustic geometry. In this work, we accomplish this task by 
studying the dependence of the value of $\kappa$ on the geometric configurations of 
the background matter flow as well as on various astrophysically relevant 
initial boundary conditions governing such flow described by different thermodynamic 
equations of state. In this way we intend
to provide a reference space spanned by fundamental accretion parameters to 
apprehend under which astrophysically relevant scenario the analogue 
effects will be pronounced. 

As already mentioned in the introduction, we 
particularly emphasize on the geometric (sonic manifold) aspects of the system 
rather than the study of the field theoretic aspects and hence the 
analysis of the origin and the properties of the Hawking like radiation/temperature 
is beyond the scope of our present work. 

We found that the Hawking like effects become more pronounced in the relatively
stronger gravity region in the sense that irrespective of the equation of state as well as the
initial boundary conditions, the acoustic surface gravity for stationary
mono-transonic solutions assumes its maximum value when the acoustic horizons
are formed at very close proximity of the black hole event horizon. Among all
three geometric configurations of the background axisymmetric flow considered
in this work, flow in hydrostatic equilibrium along the vertical direction
produces the acoustic horizons of smallest radius and hence the corresponding
surface gravity and the Hawking like temperature becomes maximum for such flow
configuration. This is true for both the adiabatic as well as the isothermal
accretion. It has also been observed that hotter flow (adiabatic flow
parameterized by large value of ${\cal E}$ or isothermal flow parameterized by
high temperature) produces the larger value of the acoustic surface gravity since for such flow
the inner type sonic points are formed very close to the black hole event horizon. Similar
effects are observed for flow with large values of $\lambda$ and $\gamma$.
One thus concludes that relatively faster rotating hotter flows are responsible for
maximizing the analogue effects for axisymmetric background flow in Schwarzschild
metric.

As already clarified in the previous sections, non-universal features of 
Hawking like effects in dispersive media depends on the value of 
the space gradient of the background flow velocity as well as that of the 
speed of propagation of perturbation for fluid flow with 
position dependent sound speed. The aforementioned calibration space 
will also be useful to point out the relevance of certain astrophysical 
configurations to simulate the set up where such deviation can 
be maximum. This will certainly be useful to study the effect of 
gravity on the non-conventional 
classical features in Hawking like effects as is expected to be observed 
in the limit of a strong dispersion relation - no such work has been 
reported in the literature yet. 

For adiabatic as well as for isothermal flows in hydrostatic equilibrium in the 
vertical direction, the critical point and the sonic points are found to be 
non-overlapping, and an integral solution of the flow equations are needed to 
obtain the location of the acoustic surface gravity. Note that whereas for the 
adiabatic flow it is true for pseudo-Schwarzschild accretion under the influence of 
the modified potentials, isothermal accretion within such modified Newtonian 
framework does not discriminate between a critical and a sonic point \cite{bcdn12}.
It is, however, difficult to conclude anything about the universality of 
such phenomena since the corresponding expression for the flow 
thickness has been derived using a set of idealized assumptions. 
A more realistic flow thickness may be derived
by employing the non-LTE 
radiative transfer \cite{hh98,dh06} or
by taking recourse to the Grad-Shafranov equations for the MHD flow \cite{beskin97,bt05,beskin09}.

In the present work, we concentrate on the calculation of the acoustic 
surface gravity $\kappa$ for mono-transonic stationary integral accretion 
solutions constructed through the inner saddle type sonic points, and the
multi-transonic flow solutions have not been considered. This, however, is 
not a limitation of our formalism. For multi-transonic 
accretion, the flow essentially passes through the inner sonic point 
anyway. We have also demonstrated that the value of $\kappa$ evaluated at 
the inner acoustic horizon is significantly larger compared to its value
evaluated at the outer acoustic horizons. Also $\kappa$ for the inner 
acoustic horizon is far more sensitive to the initial boundary conditions.
To understand the influence of the geometric configuration of matter 
on the sonic geometry, it is thus sufficient to study the mono-transonic
 flow through the inner acoustic horizon, at least for the 
present context as described above. 

Study of the acoustic geometry for an entire shocked flow 
(with two solutions constructed through the outer and the inner 
sonic horizons connected by a discontinuous shock of practically zero 
thickness) might have other interesting consequences. The discriminant 
of the acoustic metric 
\begin{equation}
{\cal D} {\equiv} G^2_{t{\phi}}-G_{tt}G_{\phi{\phi}}
\label{discriminant}
\end{equation}
vanishes at the regular acoustic horizons. The notion of the 
sub (super) -- to super (sub) sonic transition may be 
understood through the sign change of the discriminant. 
Stationary acoustic horizons are located at the corresponding 
radial distances where ${\cal D}$ changes its sign. A regular 
smooth transition of ${\cal D} > 0 {\longrightarrow} {\cal D} <0$ 
kind represents an acoustic black hole and a discontinuity 
of
${\cal D} > 0 {\longrightarrow} {\cal D} <0$ type characterizes a shock, 
which actually can be considered as an acoustic white hole, following 
the notion introduced in \cite{blsv}. The major concern in dealing with 
a multi-transonic accretion is, however, the divergence of 
$\left[u_{\perp},c_s\right]$ (and hence of their space gradients 
$\left[\frac{du_{\perp}}{dr},\frac{c_s}{dr}\right]$) at the 
shock location. Since 
\begin{equation} 
\kappa{\equiv}\kappa\left[u_{\perp},c_s,\frac{du_{\perp}}{dr},\frac{dc_s}{dr}\right],
\label{surface-gravity-dependence-on-velocity-variables}
\end{equation}
the acoustic surface gravity also diverges in this case at the shock. If the 
analogue temperature is assumed to be directly proportional to the acoustic 
surface gravity, then such temperature can not be evaluated at shock 
as it seems. This apparently seems to be true for any discontinuity 
similar to the first order phase transition in general (shock transition 
considered in this paper shares the properties of a first order phase 
transition in thermodynamics). This is in accordance with the 
general remarks made in \cite{leonhardt-hawking-radiation-dispersive-media,
hawking-radiation-in-dispersive-media-review-article} where the formalism for obtaining 
the Hawking like temperature fails to provide acceptable results 
for a considerably large value of the space gradient of the background flow velocity 
normal to the acoustic horizon ($u_{\perp}$ -- which is the advective velocity 
for the flow configurations considered in the present work). 

There are, however, ample scopes for further critical discussion 
in this aspect. In the present work, irrotational inviscid flow 
with infinitely thin shock surface has been considered, leading to
a sharp discontinuity in $\left[u_{\perp},c_s\right]$ at the shock. For 
large scale relativistic astrophysical flows, however, the effect
of viscosity and cosmic magnetic field may not be underestimated, resulting 
in the presence of various dissipative mechanisms (bulk heating and cooling for 
background flow, even if the steady flow remains a stable one) through 
the Comptonization, bremstruhlung and synchrotron mechanism 
\cite{rybicki-lightman,kat98}. 
It may even be subjected to the turbulent instabilities as well, and non linear perturbation may
have non-negligible effects in studying the associated acoustic geometry. Effects of the
Velikhov-Chandrasekhar \cite{chandrasekhar-instability} or the Balbus-Hawley \cite{babus-hawley} type
magneto rotational instability (MRI) may become significant close to the gravitational horizon.
Clearly, such flow can be considered neither inviscid nor irrotational, and will not 
allow the formation of zero-thickness standing shock -- shock may still form 
but it will have finite thickness \cite{landau} with practically continuously (albeit 
quite steeply) varying flow variables (and their respective space gradients) at the shock.
As a result, the discontinuity of $\left[u_{\perp},c_s,\frac{du_{\perp}}{dr},\frac{dc_s}{dr}\right]$
may be smeared out. However, they can possess large values resulting in a very large value of
$\kappa$ which can be considered as an astrophysical manifestation of the general result 
reported in \cite{lib00}. 

For high viscous flow, even for steady hydrodynamic accretion, shock 
may not even form as an extreme case. That excludes the possibility of 
obtaining the aforementioned large value of $\kappa$ for such background configuration. 
For shock-free multi-transonic flow, the stationary solutions 
constructed through the inner and the outer saddle type sonic
points may join through a middle spiral type (instead 
of a centre type) critical point. 

All such additional complexities (in the background flow 
configuration) mentioned above may bring the system into the 
risk of destroying the Lorentz invariance. Analytical modelling of
such analogue systems may not be amenable. One has to take recourse to the 
large-scale numerical simulation to understand the dynamics of such a 
complex fluid configuration. 
From the analogue point of view, viscosity is likely to
destroy the Lorentz invariance,
and the assumptions behind constructing an acoustic geometry may not
be quite consistent for such a case as mentioned above.

In our work, therefore, we 
consider only the inviscid flow. 
One of the significant effects of inclusion of the viscosity
would be the
reduction of the angular momentum. As we demonstrate in this work, the
location of the acoustic horizon anti-correlates with $\lambda$.
Weakly rotating
flow makes the dynamical velocity gradient steeper leading to the
conclusion that
for viscous flow the acoustic horizons will be pushed further out from the
black hole and the flow would become supersonic at a larger distance for the
same set of other initial boundary conditions. The value of the acoustic surface
gravity anti-correlates with the location of
the acoustic horizon. A viscous transonic accretion disc is thus expected to
produce lower value of $\kappa$ compared to its inviscid counterpart.

As is obvious, we concentrate on stationary background solutions. 
Since transient phenomena are not quite uncommon in large scale 
accretion processes, it is thus important to
ensure the stability of such stationary solutions - at least within a reasonable
astrophysical time scale. One of our recent works 
\cite{deepika-schwarzschild} 
accomplished such task by perturbing the
corresponding space-time dependent fluid dynamic equations governing the
accretion process and by studying whether such perturbation converges to
ensure the stability of the transonic solutions of the time
independent part of the aforementioned fluid dynamic equations.
Work presented in \cite{deepika-schwarzschild}
ensures that the corresponding stationary
integral transonic solutions are stable and demonstrates that the relativistic
acoustic geometry emerges from such perturbation analysis, as well as it 
demonstrates that the acoustic metric obtained for such configurations is 
independent of the mode of perturbation, i.e., the sonic metric 
as well as $\kappa$ remains invariant irrespective of whether the 
velocity potential or the mass accretion rate is perturbed
using the linear perturbation technique. Such works may be 
considered as the dynamical treatment (to study the sonic 
geometry) corresponding to the stationary approach presented in 
our present work. In such works too, however, the shock related 
discontinuity can not be avoided (if shock forms in inviscid fluid)
since the entire stability analysis scheme had been based on the linear perturbation 
analysis (where sound wave is defined as a small amplitude linear perturbation, 
following the general convention). 

There are, however, recently published works on the possibility of 
calculation of the Hawking like temperature for flow with discontinuities. 
For dilute Bose-Einstein condensates with a step like horizon, 
microscopic Bogoliubov theory has been applied \cite{rpc}
to obtain a closed form analytical expression for the spectral distribution 
of the analogue radiation, even if the corresponding surface gravity 
diverges because of the discontinuous profile. Expression for the 
Hawking like temperature could not be obtained but the 
spectrum had been predicted to be a thermal one for such case. 
Later on, a series of works by Finazzi and his collaborators 
\cite{finazzi1,finazzi2} used non-linear dispersion relation 
to obtain the analogue temperature for Bose-Einstein condensates
(and claimed that the result holds good for more general kind 
of flow as well), even if the profile contains 
discontinuity. In their work, a critical length had been defined as a function of 
the healing length $\zeta$ (originally defined by \cite{unr95} 
and later on re-interpreted by \cite{cpf} in the relevant 
context) and the acoustic surface gravity $\kappa$. The gradient of the 
dynamical velocity was then computed over the aforementioned critical length scale 
across the horizon to obtain the emergent spectrum. Even if the analogue temperature 
differs considerably from the Hawking one, the corresponding flux still comes 
out to be thermal. 

We are, however, not sure whether the analogue temperature can be found 
out at the shock location for the flow profile we consider. Being an 
inviscid flow, the shock is infinitely thin, and the discontinuity 
resembles a delta function, which rules out the possibility of any 
`smooth averaging' over such discontinuity (had it been the case that 
the shock would have finite thickness, such an averaging procedure 
could perhaps be relevant in the present context). Also our perturbation 
analysis is linear and no leading term with higher order perturbation 
is considered while obtaining the corresponding wave equation 
(for the propagation of the perturbation) and the associated acoustic metric 
element $G_{\mu\nu}$. Hence application of the methodology developed 
in \cite{finazzi1,finazzi2} in our work to compute the corresponding 
analogue temperature and the associated spectra at the shock would be 
considerably involved and is, unfortunately, beyond the scope of the
present paper. Nevertheless, it will really be interesting to see 
whether such a methodology could be adopted for the shocked flow 
configuration considered here since it would open up an unique 
possibility of obtaining very large (and hence, observationally detectable)
analogue temperature at the shock. We plan to perform such task 
in future for astrophysical flows with finite shock thickness. 

Our present work deals with accretion onto non-rotating black holes. 
The hypothesis that some of the supermassive black holes and the
stellar mass black holes powering the active galactic nuclei
and the galactic microquasars, respectively, possess non-zero
values of the spin angular momentum in reality, has gained widespread acceptance in recent times
\cite{mil09,kmtnm10,zio10,tch10,daly11,bggnps11,rey11,mcc11,mar11,dau10,nix11,tch12,
McKinney-Tchekhovskoy-Blandford,Brenneman,Dotti-Colpi-Pallini-Perego-Volonteri,
Sesana-Barausse-Dotti-Rossi,Fabian-Parker-Wilkins-Miller-Kara-Reynolds-Dauser,
Healy-Lousto-Zlochower,Jiang-Bambi-Steiner,Nemmen-Tchekhovskoy}.
The black hole spin
plays a deterministic role in influencing
various characteristic dynamical and spectral features of accretion and
related phenomena in the characteristic metric. We 
thus feel that it is imperative to study the influence of 
the black hole spin angular momentum on the emergence of the 
relativistic sonic geometry for black hole accretion 
in the Kerr metric. In our next work we shall report 
how the $\kappa - a$ ($a$ being the Kerr parameter)
profile differs for various flow geometries and 
equations of state. This will greatly help 
to understand how the actual black hole metric 
influences the properties of the emergent 
acoustic metric. In addition, from 
recent theoretical and observational findings, the
relevance of the counter-rotating accretion in black hole
astrophysics is being increasingly evident
\cite{dau10,nix11,tch12,garofalo2013}.
It is thus instructive to study whether the characteristic features
of the various $\kappa - a$ profiles manifest any discrimination 
between a prograde and a retrograde background flow. 

\section{Appendices}
\label{apen}
\subsection{Determination of the location of the sonic horizon(s) and the 
critical velocity gradients at such horizons -- Polytropic accretion. }
\label{apen-poly}
In this section we shall demonstrate how one can obtain an algebraic polynomial 
in $r_c$, the critical point of the flow, and how the computation of the
location of the acoustic horizons $r_h$ follows from such procedure. We then 
obtain the values of the dynamical velocity (radial advective velocity on 
the equatorial plane) and the sonic velocity (speed of propagation of the 
embedded linear perturbation), as well as their space gradients, evaluated 
on the acoustic horizon $r_h$. This completes the exact estimation of the 
value of the acoustic surface gravity $\kappa$ since 
\begin{equation} 
\kappa{\equiv}\kappa\left[u,c_s,\frac{du}{dr},\frac{dc_s}{dr}\right]_{r_h}
\label{appen1}
\end{equation} 
as has been shown in eq. (\ref{anal4}). \\ \\
\noindent
The total specific energy first integral of motion 
\noindent
\begin{equation}
{\cal E} = \frac{\gamma -1}
{(\gamma -(1+c_s{}^2))}\sqrt{\frac{(1-\frac{2}{r})}{(1-\frac{\lambda ^2}{r^2}(1-\frac{2}{r}))(1-u^2)}}
\label{apen2}
\end{equation}
remains conserved for every $r$, including $r_c$ (and $r_h$, if $r_h{\ne}r_c$, we shall discuss 
such cases in detail), for polytropic accretion. 
\noindent
For the constant height flow, 
critical point condition comes out to be (see eq. \ref{anal49})
\begin{equation}
\left[u=c_s\right]_{r_c}=
\sqrt{\frac{f_2(r_c,\lambda )}{\frac{1}{r_c}+\frac{1}{r_c{}^2(1-\frac{2}{r_c})}}}
\label{apen3}
\end{equation}
where $f_2\left(r_c,\lambda\right)$ are obtained by substituting $r=r_c$ in the expression 
provided by eq. (\ref{anal48b}). \\ \\
\noindent
Substitution of the value of $u$ and $c_s$ evaluated at the critical point
(expressed through the eq.(\ref{apen3})) in eq. (\ref{apen2}) provides the 
following 11$^{\rm th}$ degree algebraic polynomial equation in $r_c$
\begin{equation}
a_0 + a_1 r_c + a_2 r_c^2 + a_3 r_c^3 + a_4 r_c^4 + a_5 r_c^5 + a_6 r_c^6 + a_7 r_c^7 + a_8 r_c^8 + a_9 r_c^9 + a_{10} r_c^{10} + a_{11} r_c^{11}=0
\label{apen4}
\end{equation}
where, the coefficients $a_i$ are given by
\begin{eqnarray}
a_0 = 4 (\gamma -3)^2 \lambda ^6 \mathcal{E}^2 \nonumber \\
\nonumber
a_1 = -4 \left(3 \gamma ^2-16 \gamma +21\right) \lambda ^6 \mathcal{E}^2 \\
\nonumber
a_2 = \lambda ^4 \left(\left(13 \gamma ^2-62 \gamma +73\right) \lambda ^2 \mathcal{E}^2-4 (\gamma -1)^2\right) \\
\nonumber
a_3 = 2 \lambda ^4 \left(2 \left(4 (\gamma -1)^2+3 (\gamma -3) \mathcal{E}^2\right)-\left(3 \gamma ^2-13 \gamma +14\right) \lambda ^2 \mathcal{E}^2\right) \\
\nonumber
a_4 = \lambda ^4 \left(-25 (\gamma -1)^2+2 \left(\gamma ^2-17 \gamma +36\right) \mathcal{E}^2+(\gamma -2)^2 \lambda ^2 \mathcal{E}^2\right) \\
\nonumber
a_5 = \lambda ^2 \left(-4 (\gamma -1)^2-\lambda ^2 \left(\left(5 \gamma ^2-38 \gamma +59\right) \mathcal{E}^2-19 (\gamma -1)^2\right)\right) \\
\nonumber
a_6 = \lambda ^2 \left(\lambda ^2 \left(4 \left(\gamma ^2-5 \gamma +6\right) \mathcal{E}^2-7 (\gamma -1)^2\right)+\gamma ^2 \left(14-3 \mathcal{E}^2\right)+4 \gamma  \left(3 \mathcal{E}^2-7\right)+14\right) \\
\nonumber
a_7 = \lambda ^2 \left(2 \left(2 \left(2 \gamma ^2-7 \gamma +3\right) \mathcal{E}^2-9 (\gamma -1)^2\right)-\lambda ^2 \left((\gamma -2)^2 \mathcal{E}^2-(\gamma -1)^2\right)\right) \\
\nonumber
a_8 = -(\gamma -1)^2-\lambda ^2 \left(\left(7 \gamma ^2-22 \gamma +13\right) \mathcal{E}^2-10 (\gamma -1)^2\right) \\
\nonumber
a_9 = -6 \gamma +\gamma ^2 \left(-\left(\mathcal{E}^2-3\right)\right)+2 (\gamma -1) \lambda ^2 \left(-\gamma +(\gamma -2) \mathcal{E}^2+1\right)+3 \\
\nonumber
a_{10} = (\gamma -1) \left(\gamma  \left(2 \mathcal{E}^2-3\right)+3\right) \\
\nonumber
a_{11} = (\gamma -1)^2 \left(1-\mathcal{E}^2\right) \\
\label{apen5}
\end{eqnarray}
A specific choice of the initial boundary conditions governing the flow and specified by 
astrophysically relevant values of $\left[{\cal E},\lambda,\gamma\right]$ is required to 
solve the above equation for $r_c$. Eq. (\ref{apen3}) implies that the critical surfaces
are isomorphic with the sonic surfaces, and hence $r_c$ and $r_h$ is identical. This 
requires that an acceptable root (of the aforementioned polynomial)  
will be real, positive and should lie outside the gravitational horizon (located at 
$r=2$ in the system of units used in the present work). \\ \\
\noindent
For conical flow, the critical point condition
\begin{equation}
\left[u=c_s\right]_{r_c}=
\sqrt{\frac{f_2(r_c,\lambda )}{\frac{2}{r_c}+\frac{1}{r_c{}^2(1-\frac{2}{r_c})}}},
\label{apen6}
\end{equation}
implies that $r_c=r_h$, and when substituted in eq. (\ref{apen2}), provides the 
following 11$^{\rm th}$ polynomial equation 
\begin{equation}
a_0 + a_1 r_c + a_2 r_c^2 + a_3 r_c^3 + a_4 r_c^4 + a_5 r_c^5 + a_6 r_c^6 + a_7 r_c^7 + a_8 r_c^8 + a_9 r_c^9 + a_{10} r_c^{10} + a_{11} r_c^{11} = 0
\label{apen7}
\end{equation}
where the coefficients $a_i$ are found to be
\begin{eqnarray}
a_0 = -4 \mathcal{E}^2 (5-3 \gamma )^2 \lambda ^6  \nonumber \\
\nonumber
a_1 = 8 \mathcal{E}^2 \left(40-49 \gamma +15 \gamma ^2\right) \lambda ^6  \\
\nonumber
a_2 = -\lambda ^4 \left(108+401 \mathcal{E}^2 \lambda ^2+\gamma ^2 \left(108+157 \mathcal{E}^2 \lambda ^2\right)-2 \gamma  \left(108+251 \mathcal{E}^2 \lambda ^2\right)\right)  \\
\nonumber
a_3 = \lambda ^4 \left(324 (-1+\gamma )^2+\mathcal{E}^2 \left(-240+247 \lambda ^2-4 \gamma  \left(-81+79 \lambda ^2\right)+\gamma ^2 \left(-108+101 \lambda ^2\right)\right)\right)  \\
\nonumber
a_4 = \lambda ^4 \left(-387 (-1+\gamma )^2+\mathcal{E}^2 \left(564-75 \lambda ^2+\gamma ^2 \left(270-32 \lambda ^2\right)+\gamma  \left(-786+98 \lambda ^2\right)\right)\right)  \\
\nonumber
a_5 = \lambda ^2 (-108+\left(230-502 \mathcal{E}^2\right) \lambda ^2+9 \mathcal{E}^2 \lambda ^4+2 \gamma ^2 \left(-54+\left(115-126 \mathcal{E}^2\right) \lambda ^2+2 \mathcal{E}^2 \lambda ^4\right)  \\
\nonumber
a_6 = \lambda ^2 \left(-2 (-1+\gamma )^2 \left(-135+34 \lambda ^2\right)+\mathcal{E}^2 \left(-84+200 \lambda ^2+\gamma  \left(180-290 \lambda ^2\right)+\gamma ^2 \left(-81+104 \lambda ^2\right)\right)\right)  \\
\nonumber
a_7 = \lambda ^2 \left(4 (-1+\gamma )^2 \left(-63+2 \lambda ^2\right)+\mathcal{E}^2 \left(\gamma ^2 \left(153-16 \lambda ^2\right)-30 \left(-6+\lambda ^2\right)+4 \gamma  \left(-87+11 \lambda ^2\right)\right)\right)  \\
\nonumber
a_8 = -27-4 \left(-26+31 \mathcal{E}^2\right) \lambda ^2+\gamma ^2 \left(-27+\left(104-96 \mathcal{E}^2\right) \lambda ^2\right)+\gamma  \left(54+16 \left(-13+14 \mathcal{E}^2\right) \lambda ^2\right)  \\
\nonumber
a_9 = 2 \left(-(-1+\gamma )^2 \left(-27+8 \lambda ^2\right)+\mathcal{E}^2 \left(-4+14 \lambda ^2+\gamma  \left(12-24 \lambda ^2\right)+\gamma ^2 \left(-9+10 \lambda ^2\right)\right)\right)  \\
\nonumber
a_{10} = 4 (-1+\gamma ) \left(9-9 \gamma +\mathcal{E}^2 (-4+6 \gamma )\right)  \\
\nonumber
 a_{11} = -8 \left(-1+\mathcal{E}^2\right) (-1+\gamma )^2 
\label{apen8}
\end{eqnarray}
For flow in vertical equilibrium, the critical point condition 
\begin{equation}
\left[u=\sqrt{\frac{1}{1+(\frac{\gamma +1}{2})(\frac{1}{c_s{}^2})}}\right]_{r_c}
=\sqrt{\frac{f_2(r_c,\lambda )}{f_1(r_c,\lambda )+f_2(r_c,\lambda)}}
\label{apen9}
\end{equation}
indicates that the critical points and the sonic points are not located at the 
same radial distance, and the value of flow variables and their 
space gradients on the sonic 
horizons are to be obtained by integrating the flow equation using the 
critical flow variables (as well as their gradients). In eq. (\ref{apen9}), 
$f_1(r_c,\lambda)$ is obtained by putting $r=r_c$ in eq. (\ref{anal48a}). 
Substituting the values of the critical velocities (as obtained through eq. (\ref{apen9})
in eq. (\ref{apen2}), the following 8$^{\rm th}$ degree polynomial equation in $r_c$ is 
obtained
\begin{equation}
a_0 + a_1 r_c + a_2 r_c^2 + a_3 r_c^3 + a_4 r_c^4 + a_5 r_c^5 + a_6 r_c^6 + a_7 r_c^7 + a_8 r_c^8 = 0
\label{apen9a}
\end{equation}
where the coefficients $a_i$ are found to be
\begin{eqnarray}
 a_0 = 64 (-2+\gamma )^2 \mathcal{E} ^2 \lambda ^4  \nonumber \\
\nonumber
a_1 = -32 (18-19 \gamma +5 \gamma ^2) \mathcal{E} ^2 \lambda ^4  \\
\nonumber
 a_2 = 4 \lambda ^2 ((121-134 \gamma +37 \gamma ^2) \mathcal{E} ^2 \lambda ^2+60 (-1+\gamma )^2 )  \\
\nonumber
 a_3 = 4 \lambda ^2 ((-45+52 \gamma -15 \gamma ^2) \mathcal{E} ^2 \lambda ^2+4 (-34+22 \mathcal{E} ^2+\gamma  (68-37 \mathcal{E} ^2)+\gamma ^2 (-34+13 \mathcal{E} ^2)) )  \\
\nonumber
 a_4 = \lambda ^2 ((5-3 \gamma )^2 \mathcal{E} ^2 \lambda ^2-4 (-115+147 \mathcal{E} ^2+\gamma  (230-244 \mathcal{E} ^2)+\gamma ^2 (-115+89 \mathcal{E} ^2)) ) \\
\nonumber
 a_5 = -2 (-60 (-1+\gamma )^2+(86-163 \mathcal{E} ^2+\gamma ^2 (86-99 \mathcal{E} ^2)+2 \gamma  (-86+133 \mathcal{E} ^2)) \lambda ^2-60 (-1+\gamma )^2 )  \\
\nonumber
 a_6 = (-12 (-1+\gamma ) (2-5 \mathcal{E} ^2+\gamma  (-2+3 \mathcal{E} ^2)) \lambda ^2+ (-384+121 \mathcal{E} ^2+\gamma  (768-286 \mathcal{E} ^2)+\gamma ^2 (-384+169 \mathcal{E} ^2)) )  \\
\nonumber
a_7 = -12 (-1+\gamma ) (17-11 \mathcal{E} ^2+\gamma  (-17+13 \mathcal{E} ^2)) \\
\nonumber
a_8 = 36 (-1+\gamma )^2 (-1+\mathcal{E} ^2)  
\label{apen10}
\end{eqnarray}
One needs to realize that unlike the other two flow models, it is necessary but not sufficient to have a root 
for the above equation as $r_c>2$, since we need $r_h>2$ and $r_h<r_c$ always. \\ \\
\noindent
Space gradient of advective velocity for various flow models can be obtained as
\begin{equation}
\left[\left(\frac{\text{du}}{\text{dr}}\right)_{r_c}\right]_{\rm i}
=\left[-\frac{\alpha_i}{2\Gamma_i}\overset{+}{-}\frac{\sqrt{\alpha_i ^2-4\Gamma_i\beta_i }}{2\Gamma_i}\right]_{r_c}
\label{apen11}
\end{equation}
where $\left[{\rm i=1,2,3}\right]$ corresponds to the constant height flow, conical flow and flow in vertical 
equilibrium, respectively. The set of quantities $\left[\alpha,\beta,\Gamma\right]_{\rm i = 1,2,3}$ can explicitly 
be obtained as
\begin{eqnarray}
\alpha_1 = \frac{2 c \left(\gamma -1-c^2\right) \left(r_c-1\right)}{\left(1-c^2\right) r_c \left(r_c-2\right)},\\
\nonumber
\beta_1 = \frac{\beta'}{(-2+r_c)^2 r_c^2 \left(r_c^3-(-2+r_c) \lambda ^2\right)^2},\\
\nonumber
\Gamma_1 = \frac{\gamma -3u_c{}^2+1}{\left(1-u_c{}^2\right){}^2}
\label{apen12}
\end{eqnarray}
where 
$\beta'=[\gamma  \lambda ^4 c_c{}^2 r_c^4-6 \gamma  \lambda ^4 c_c{}^2 r_c^3+13 \gamma  \lambda ^4 c_c{}^2 r_c^2-12 \gamma  \lambda ^4 c_c{}^2 r_c+4 \gamma  \lambda ^4 c_c{}^2-2 \gamma  \lambda ^2 c_c{}^2 r_c^6+8 \gamma  \lambda ^2 c_c{}^2 r_c^5-10 \gamma  \lambda ^2 c_c{}^2 r_c^4+4 \gamma  \lambda ^2 c_c{}^2 r_c^3+\gamma  c_c{}^2 r_c^8-2 \gamma  c_c{}^2 r_c^7+\gamma  c_c{}^2 r_c^6-\lambda ^4 r_c^4-\lambda ^4 c_c{}^4 r_c^4+8 \lambda ^4 r_c^3+6 \lambda ^4 c_c{}^4 r_c^3-24 \lambda ^4 r_c^2-13 \lambda ^4 c_c{}^4 r_c^2+\lambda ^4 c_c{}^2 r_c^2+32 \lambda ^4 r_c+12 \lambda ^4 c_c{}^4 r_c-4 \lambda ^4 c_c{}^2 r_c-4 \lambda ^4 c_c{}^4+4 \lambda ^4 c_c{}^2+2 \lambda ^2 c_c{}^4 r_c^6+3 \lambda ^2 r_c^6-8 \lambda ^2 c_c{}^4 r_c^5-20 \lambda ^2 r_c^5+10 \lambda ^2 c_c{}^4 r_c^4+48 \lambda ^2 r_c^4-2 \lambda ^2 c_c{}^2 r_c^4-4 \lambda ^2 c_c{}^4 r_c^3-40 \lambda ^2 r_c^3+4 \lambda ^2 c_c{}^2 r_c^3-c_c{}^4 r_c^8+2 c_c{}^4 r_c^7-2 r_c^7-c_c{}^4 r_c^6+c_c{}^2 r_c^6+2 r_c^6-16 \lambda ^4]_{r_c}$
\begin{eqnarray}
\alpha_2 = \frac{2 c \left(\gamma -1-c^2\right) \left(2 r_c-3\right)}{\left(1-c^2\right) r_c \left(r_c-2\right)},\\
\nonumber
\beta_2 = \frac{\beta''}{(-2+r)^2 r^2 \left(r^3-(-2+r) \lambda ^2\right)^2},\\
\nonumber
\Gamma_2 = \frac{\gamma -3u_c{}^2+1}{\left(1-u_c{}^2\right){}^2}
\label{apen14}
\end{eqnarray}
where
$\beta''=[-36 c^4 \lambda ^4-4 c^4 \lambda ^4 r_c^4+28 c^4 \lambda ^4 r_c^3-73 c^4 \lambda ^4 r_c^2+84 c^4 \lambda ^4 r_c+8 c^4 \lambda ^2 r_c^6-40 c^4 \lambda ^2 r_c^5+66 c^4 \lambda ^2 r_c^4-36 c^4 \lambda ^2 r_c^3-4 c^4 r_c^8+12 c^4 r_c^7-9 c^4 r_c^6+36 c^2 \gamma  \lambda ^4-12 c^2 \lambda ^4+4 c^2 \gamma  \lambda ^4 r_c^4-28 c^2 \gamma  \lambda ^4 r_c^3+73 c^2 \gamma  \lambda ^4 r_c^2-84 c^2 \gamma  \lambda ^4 r_c-8 c^2 \gamma  \lambda ^2 r_c^6+40 c^2 \gamma  \lambda ^2 r_c^5-66 c^2 \gamma  \lambda ^2 r_c^4+36 c^2 \gamma  \lambda ^2 r_c^3+4 c^2 \gamma  r_c^8-12 c^2 \gamma  r_c^7+9 c^2 \gamma  r_c^6-2 c^2 \lambda ^4 r_c^4+14 c^2 \lambda ^4 r_c^3-35 c^2 \lambda ^4 r_c^2+36 c^2 \lambda ^4 r_c+4 c^2 \lambda ^2 r_c^6-20 c^2 \lambda ^2 r_c^5+30 c^2 \lambda ^2 r_c^4-12 c^2 \lambda ^2 r_c^3-2 c^2 r_c^8+6 c^2 r_c^7-3 c^2 r_c^6-\lambda ^4 r_c^4+8 \lambda ^4 r_c^3-24 \lambda ^4 r_c^2+32 \lambda ^4 r_c+3 \lambda ^2 r_c^6-20 \lambda ^2 r_c^5+48 \lambda ^2 r_c^4-40 \lambda ^2 r_c^3-2 r_c^7+2 r_c^6-16 \lambda ^4]_{r_c}
$
\begin{eqnarray}
\alpha_3 = \frac{8 c_c^2 \left(\gamma -1-c_c^2\right) \left(-2 \lambda ^2 r_c+3 r_c^3+3 \lambda ^2\right)}{(\gamma +1)^2 r_c u_c \left(-\lambda ^2 r_c+r_c^3+2 \lambda ^2\right)}, \nonumber \\
\beta_3 = -\left[\frac{\beta_3''}{(\gamma +1)^2 \left(r_c-2\right){}^2 r_c^2 \left(-\lambda ^2 r_c+r_c^3+2 \lambda ^2\right){}^2}\right]_c,
\Gamma_3 = -\frac{4 c_c{}^4}{(\gamma +1)^2 u_c{}^2}+\frac{2 (3 \gamma -1) c_c{}^2}{(\gamma +1)^2 u_c{}^2}+\frac{u_c{}^2+1}{\left(u_c{}^2-1\right){}^2} \nonumber \\
\label{apen16}
\end{eqnarray}
The critical gradient of sonic velocities can be obtained by putting the values of 
$\left[\left(\frac{du}{dr}\right)_{\rm r_c}\right]_{\rm i}$ at eq. (\ref{anal46}, \ref{anal59},
\ref{anal67}) for ${\rm i = 1,2,3}$, respectively. For the constant height flow as well 
as the conical flow, $\left[u,c_s,\frac{du}{dr},\frac{dc_s}{dr}\right]_{\rm r_c}$ thus obtained 
can directly be substituted in eq. (\ref{anal17}) to obtain the value of the 
acoustic surface gravity. For flow in hydrostatic equilibrium (along the vertical 
direction), $\left[u,c_s,\frac{du}{dr},\frac{dc_s}{dr}\right]_{\rm r_h}$ has to be 
obtained by integrating the flow equations using the the corresponding values of the 
$\left[u,c_s,\frac{du}{dr},\frac{dc_s}{dr}\right]_{\rm r_c}$, and then by 
substituting $\left[u,c_s,\frac{du}{dr},\frac{dc_s}{dr}\right]_{\rm r_h}$ in 
eq. (\ref{anal17}). 

\subsection{Determination of the location of the sonic horizon(s) and the
critical velocity gradients at such horizons -- Isothermal accretion.}
\label{apen-iso}
\noindent
Here, the first integral of motion of the following form (see, e.g., eq. (\ref{anal74}--\ref{anal73}))
\begin{equation}
\xi=\frac{r^2(r-2)}{(r^3-(r-2) \lambda ^2) (1-u^2)} \rho^{{\frac{2k_B}{\mu m_H} T}}
\label{apen18}
\end{equation}
serves as the equivalent of the energy first integral of motion in adiabatic flow \footnote{For 
isothermal accretion total energy is not conserved.} in connection with the computation 
of location of the critical points. The critical point conditions for the 
constant height flow is found to be 
\begin{equation}
\left[u^2=c_s^2=\frac{- r^3+ (r-2)^2 \lambda ^2}{r^3-r^4+(r-2)(r-1)\lambda ^2}\right]_{r_c}
\label{apen19}
\end{equation}
Substituting the expressions for the critical velocity and the critical sound speed 
(as obtained in the above equation) in eq. (\ref{apen18}), we obtain the following 
$4^{\rm th}$ degree polynomial equation in $r_c$ 
\begin{equation}
\left[2 c_s^2 r^4 - 2 \left(1+c_s^2\right)r^3 - 2\lambda ^2\left(c_s^2-1\right)r^2 - 
2\text{$\lambda $}^2\left(4-3c_s^2\right)r - 4 \lambda ^2\left(c_s^2-2\right)\right]_{r_c} = 0
\label{apen20}
\end{equation}
The critical point condition and the corresponding polynomial equation for the 
conical flow can be obtained as
\begin{equation}
\left[u^2=c_s^2=\frac{- r^3+ (r-2)^2 \lambda ^2}{3 r^3-2 r^4+6 \lambda ^2-7 r \lambda ^2+2 r^2 \lambda ^2} \right]_{r_c}
\label{apen21}
\end{equation}
\begin{equation}
\left[4 c_s^2 r^4-2\left(3c_s^2+1\right)r^3-2\lambda ^2\left(2c_s^2-1\right)r^2+2\text{$\lambda $}^2\left(7c_s^2-4\right)r
-4\lambda ^2\left(3c_s^2-2\right)\right]_{r_c} = 0
\label{apen22}
\end{equation}
For both of the above flow configurations, $r_c$ coincides with $r_h$ as is obvious from the 
corresponding critical point conditions. 
For flow in vertical equilibrium, $r_c{\ne}r_h$ as is evident from the following critical point condition. 
\begin{equation}
\left[u^2=\frac{c_s^2}{1+c_s^2}=\frac{- r^3+(r-2)^2 \lambda ^2}{8 r^3-4 r^4+10 \lambda ^2-11 r \lambda ^2+3 r^2 \lambda ^2}\right]_{r_c}
\label{apen23}
\end{equation}
So for the following polynomial equation which is to be solved to obtain the critical points, a necessary but 
not sufficient condition of physically acceptable roots is $r_c>2$, and $r_h$ has to 
lie outside the event horizon. 
\begin{equation}
\left[4 r_c^4 c_s^2 - r^3 \left(1+8\text{c}_s^2\right) - r^2 \lambda ^2 \left(-1+3c_s^2\right) -
{r\lambda}^2\left(4-11c_s^2\right) - 2\lambda ^2\left(-2+5c_s^2\right)\right]_{r_c} = 0
\label{apen24}
\end{equation}
The corresponding critical gradient of the advective velocity for three flow models can be 
expressed as
\begin{equation}
\left[\left(\frac{\text{du}}{\text{dr}}\right)_{r_c}\right]_{\rm CH}^{\rm Isothermal}=
\left[\frac{\alpha_1^{\rm iso}}{2\Gamma_1^{\rm iso}}\overset{+}{-}\frac{\sqrt{{\alpha_1^{iso}}^2+
4\beta_1^{\rm iso}\Gamma_1^{iso}}}{2\Gamma_1^{\rm iso} }\right]_{r_c}
\label{apen24a}
\end{equation}
\begin{equation}
\left[\left(\frac{\text{du}}{\text{dr}}\right)_{r_c}\right]_{\rm CF}^{\rm Isothermal}=
\left[\frac{\alpha_2^{\rm iso}}{2\Gamma_1^{\rm iso}}\overset{+}{-}\frac{\sqrt{{\alpha_2^{iso}}^2+
4\beta_2^{\rm iso}\Gamma_1^{iso}}}{2\Gamma_1^{\rm iso} }\right]_{r_c}
\label{apen24b}
\end{equation}
\begin{equation}
\left[\left(\frac{\text{du}}{\text{dr}}\right)_{r_c}\right]_{\rm VE}^{\rm Isothermal}=
\left[\frac{\alpha_3^{\rm iso}}{2\Gamma_3^{\rm iso}}\overset{+}{-}\frac{\sqrt{{\alpha_3^{\rm iso}}^2
+4\beta_3^{\rm iso}\Gamma_3^{\rm iso}}}{2\Gamma_3^{\rm iso} }\right]_{r_c}
\label{apen24c}
\end{equation}
The set of quantities $\left[\alpha^{iso},\beta^{iso},\Gamma^{iso}\right]_{\rm i}$ can explicitly
be obtained as
\begin{eqnarray}
\alpha_1^{\rm iso} = -\left(3 c_s{}^2-1\right)\left\{\left(-1+c_s{}^2\left(r_c-1\right)\right)r_c{}^3-\left(2+c_s{}^2\left(r_c-1\right)-r_c\right)\left(r_c-2\right)\lambda ^2\right\} \nonumber \\
\nonumber
\alpha_2^{\rm iso} = -\left(3 c_s{}^2-1\right)\left\{\left(-1+c_s{}^2\left(2r_c-3\right)\right)r^3-
\left(2+c_s{}^2\left(2r_c-3\right)-r_c\right)\left(r_c-2\right)\lambda ^2\right\} \\
\nonumber
\alpha_3^{\rm iso}=
-\left\{\left(-1+4c_s{}^2\left(r_c-2\right)\right)r_c{}^3-\left(2+c_s{}^2\left(3r_c-5\right)-r_c\right)\left(r_c-2\right)\lambda ^2\right\}\left(\frac{3c_s{}^2}{1+c_s{}^2}-1\right)\\
\beta_1^{\rm iso}= c_s\left(1-c_s{}^2\right)\left\{\left(-3+c_s{}^2\left(4r_c-3\right)\right)r_c{}^2+\left(c_s{}^2\left(3-2r_c\right)+
2\left(r_c-2\right)\right)\lambda ^2\right\} \nonumber \\
\nonumber
\beta_2^{\rm iso}= c_s\left(1-c_s{}^2\right)\left\{\left(-3+c_s{}^2\left(8r_c-9\right)\right)r_c^2+\left(c_s{}^2\left(7-4r_c\right)+
2\left(r_c-2\right)\right)\lambda ^2\right\}\\
\nonumber
\beta_3^{\rm iso}=
\left\{\left(-3+8c_s{}^2\left(2r_c-3\right)\right)r_c{}^2+\left(c_s{}^2\left(11-6r_c\right)+2\left(r_c-2\right)\right)\lambda ^2\right\}\left\{\frac{c_s}{\left(1+c_s{}^2\right){}^{\frac{3}{2}}}\right\}\\
\nonumber
\Gamma_1^{\rm iso}=2 r_c \left(r_c-2\right) \left(r_c^3 - \left(r_c-2\right) \lambda ^2\right) c_s\\
\nonumber
\Gamma_3^{\rm iso}=
2 r_c \left(r_c-2\right) \left(r_c{}^3 - \left(r_c-2\right) \lambda ^2\right) c_s\sqrt{1+c_s{}^2}
\label{apen26}
\end{eqnarray}
Isothermal flow has position independent sound speed for obvious reason. 
Note that for {\it all} flow models for the isothermal accretion, the location of the 
critical points can be found completely analytically. For constant height flow as well as 
for the conical flow model, $\left[u,c_s,\frac{du}{dr}\right]_{\rm r_c}$ thus obtained
can directly be substituted in eq.(\ref{anal17}), with a modification that 
$\frac{dc_s}{dr}=0$, to obtain the value of the
acoustic surface gravity. For flow in hydrostatic equilibrium (along the vertical
direction), $\left[u,c_s,\frac{du}{dr}\right]_{\rm r_h}$ has to be
obtained by integrating the flow equations using the corresponding values of the
$\left[u,c_s,\frac{du}{dr}\right]_{\rm r_c}$, and then by
substituting $\left[u,c_s,\frac{du}{dr}\right]_{\rm r_h}$ in
eq.(\ref{anal17}), $\kappa$ can be evaluated.

\section*{Acknowledgements}
\noindent
PT would like to acknowledge the kind hospitality
provided by HRI, Allahabad, India, for several visits. The work of TKD has been partially supported by a research grant
provided by S. N. Bose National Centre for Basic
Sciences, Kolkata, India, under a guest scientist (long term
sabbatical visiting professor) research
programme.
The authors acknowledge insightful discussions with Archan S Majumdar
and would like to sincerely thank the anonymous referee for suggesting a
number of useful improvements.
Authors also acknowledge the help of Reema Chowdhury
in correcting certain typos in the manuscript.

\end{document}